# Effects of the neutral dynamics model on the particle-in-cell simulations of a Hall thruster plasma discharge

F. Faraji [*†], M. Reza[*], A. Knoll[*]

[*]Plasma Propulsion Laboratory, Department of Aeronautics, Imperial College London, Exhibition Road, London, SW7 2AZ, United Kingdom

[†]Corresponding Author (f.faraji20@imperial.ac.uk)

**Abstract**: The dynamics of the neutral atoms in Hall thrusters affect several plasma processes, from the ionization to the electrons' mobility. In the context of Hall thruster's particle-in-cell (PIC) modeling, the neutrals are often treated kinetically, similar to the plasma species, and their interactions with themselves and the ions are resolved using the Direct-Simulation Monte-Carlo (DSMC) algorithm. However, the DSMC approach is computationally resource demanding. Therefore, modeling the neutrals as a 1D fluid has been also pursued in simulations that do not involve the radial coordinate and, hence, do not resolve the neutrals' radial expansion. In this article, we present an extensive study on the sensitivity of the PIC simulations of Hall thruster discharge to the model used for the neutral dynamics. We carried out 1D axial PIC simulations with various fluid and kinetic models of the neutrals as well as self-consistent quasi-2D axial-azimuthal simulations with different neutrals' fluid descriptions. Our results show that the predictions of the simulations in either 1D or 2D configurations are highly sensitive to the neutrals' model, and that different treatments of the neutrals change the spatiotemporal evolution of the discharge. Moreover, we observed that considering the ion-neutral collisions causes a significant variation in the neutral temperature, thus requiring that the neutrals' energy equation to be included as well in their fluid system of equations. Finally, the self-consistent axial-azimuthal simulations highlighted that a neutrals' model based on the continuity conservation equation only is not an appropriate choice and leads to physically unexpected high-frequency global discharge oscillations.

## Section 1: Introduction

Hall thrusters are electrostatic in-space plasma propulsion devices which produce thrust through the ionization of a neutral propellant, typically xenon or krypton, and the acceleration of the generated plasma ions. The plasma in Hall thrusters is immersed in a cross electric and magnetic field configuration. The strength of the externally applied magnetic field, commonly having an almost radial distribution, is such that the plasma electrons are magnetized and undergo an E × B drift motion around the thruster's discharge channel along the azimuthal direction whereas the ions are almost unmagnetized and only experience the mostly axial self-sustained electric field. The high-performance operation of Hall thrusters relies, on the one hand, on the ability to highly ionize the neutral propellant that is constantly being injected into the channel and, on the other hand, on the effective hinderance of the cross-magnetic-field motion of the electrons, which in turn determines the strength of the accelerating axial electric field.

In this respect, the dynamics of the neutral atoms, characterized by a relatively low-frequency temporal evolution and a large spatial gradient along the thruster's channel axis, play a significant role in the performance and stability of Hall devices' operation. Indeed, the neutrals can influence both the ionization and the acceleration processes in Hall thrusters. Concerning the former, the periodic depletion of the neutrals due to the ionization and their replenishment as a result of the propellant injection is believed to trigger the so-called "breathing mode" oscillations [1][2]. This characteristic oscillation mode is one the most prominent influences of the neutrals on the global response of the plasma in Hall thrusters. Moreover, the axial drift velocity of the neutrals, which determines the residence time of the particles within the channel, is shown to influence the ionization process and the overall stability of the plasma discharge [3].

Regarding the acceleration process, the collisional interactions between the neutral atoms and the plasma electrons provide the classical means of electrons' axial transport across the magnetic field, which counteracts to some extent the impedance imposed by the magnetic field against the axial motion of the electrons. More importantly, electron-neutral collisions are demonstrated to act as a regulatory mechanism for the excitation of high-frequency microscopic azimuthal instabilities within each cycle of the discharge evolution [4]. These high-frequency instabilities are shown to induce a notable cross-field electron transport [2][5], thus, affecting, in part, the effectiveness of the acceleration process.

Accordingly, proper modeling of the neutral dynamics is important for an accurate prediction of the plasma discharge behavior in Hall thrusters from the simulation codes. In this regard, two approaches are typically pursued in plasma simulations [6]: (1) the Langragian (kinetic) approach in which the neutrals are samples of a distribution function, or macroparticles, and their evolution is resolved by integrating the equations of motion for each macroparticle, (2) the fluid approach in which the neutral atoms are treated as a monoenergetic flow. In the former treatment, the collisional interactions between the neutrals themselves and between the neutral atoms and the plasma ion species is modeled following the Direct-Simulation Monte-Carlo collisions algorithm [7][8]. The kinetic-DSMC treatment of the neutrals has been used in kinetic particle-in-cell (PIC) [9][10] and hybrid fluid/PIC [11][12] simulations. The neutrals' fluid modeling has been pursued across different plasma simulation methodologies, from full-fluid to PIC [13]-[15]. Particularly in the case of PIC simulations, the fluid description of the neutrals is proposed as a solution to lower the computational cost of these plasma models [15]. This approach can be especially helpful for simulations that do not involve the radial coordinate and whose principal purpose is to investigate the axial dynamics and the behavior of the plasma discharge over extended simulation times. It is noted that the DMSC algorithm used to handle the neutrals' collisions in their kinetic description is inherently computationally burdensome as it involves particle sorting and requires a notable number of floating-point operations. As such, it can be most suited for detailed studies of the initial discharge transient (on the order of a few tens of microsecond simulation time) where the spatial distribution of the neutrals is expected to play a critical role.

When the neutrals are described as a flow in PIC simulations, the typical approach is to only solve the continuity conservation equation to resolve the temporal evolution of the neutral density [4][16]. This is also the usual approach followed in most full-fluid simulations and in the hybrid codes that treat the neutrals as a fluid [6]. Most recently, it has been proposed to solve both the continuity and the momentum conservation equations for the neutrals' flow in the 2D axial-azimuthal PIC simulations of Hall thrusters to capture the variations in both the neutral density and velocity [15][17].

Nonetheless, there has been only one very recent study on the impact that using various fluid/kinetic models of the neutrals can have on the predictions of the PIC simulations. Indeed, in Ref. [18], the authors investigated the effect of neutrals' model on the *axial-radial* PIC simulations and showed that the dynamics of the discharge is notably affected by the neutrals' model. The complementarity of the research questions and the extended methodologies of the present article with respect to this recent publication underlines the timeliness and the significance of this effort. This is specially the case since the studies in Ref. [18] were carried out using rather simple neutrals' models with simulations in which the physical constants were scaled for speed-up purposes. Additionally, despite the ion-neutral collisions have important effects on the neutrals' dynamics, particularly in the near-plume of the Hall thrusters [19], their effects have not yet been incorporated in the fluid models of the neutrals in the PIC codes. Hence, the influence of the ion-neutral momentum transfer on the neutrals' evolution in their fluid description is unknown. Finally, the recombination of the ions on the anode of Hall thrusters has been recently shown to play a role in the onset and the behavior of the breathing mode oscillations [20][21]. However, the significance of this phenomenon from the PIC simulations when implementing various fluid models of the neutral atoms is also not studied.

Motivated by the above, we have performed 1D axial and reduced-order quasi-2D (Q2D) axial-azimuthal PIC simulations of a setup representative of the SPT100 Hall thruster using different fluid models of the neutrals. The 1D axial simulations are also carried out using the kinetic-DSMC description of the neutral atoms. The implementation of the neutrals' kinetic and fluid models in our simulations is described in detail in this paper. The numerical studies of this work are performed using two of the particle-in-cell codes developed at Imperial Plasma Propulsion Laboratory (IPPL). For the 1D simulations, we have used IPPL-1D code [22], whereas the Q2D simulations are carried out using the novel IPPL-Q2D code [23]. IPPL-1D is verified against the well-known CCP benchmark case [22]. Moreover, the in-depth verification results of the IPPL-Q2D in the axial-azimuthal and the radial-azimuthal configurations can be found, respectively, in Ref. [23] and [24].

The main objectives of this article are as follows: (1) to study the sensitivity of the results of kinetic simulations to the model used for the evolution of neutral species, (2) to investigate the extent to which a fluid description of the neutral particles, in the absence and the presence of the ion-neutral collisions, can be representative of their kinetic-DSMC treatment in terms of the predicted dynamics of the plasma and the neutrals in axial and axial-azimuthal simulations, (3) to assess the impact of the anode ion recombination on the predictions of the simulations with various fluid models of the neutrals, and (4) to evaluate the influence of the neutrals' fluid models on the results of simulations with self-consistent resolution of the electrons' mobility. A supporting parametric study to understand the influence of the electron-wall collisions model on the simulations is also performed in the 1D configuration and using a neutrals' fluid model based on the continuity conservation equation.

## Section 2: Overview of the simulations' setup and conditions

The setup of the simulations is, overall, based on a test case defined within the LANDMARK benchmarking project [25] for the verification of the 1D axial fluid and hybrid fluid/PIC codes. As a result, the simulations presented in this work extend the LANDMARK's 1D benchmark case to comprise the kinetic 1D axial as well as the Q2D axial-azimuthal simulations.

The domain of the Q2D simulations is a Cartesian $x-y$ plane, representative of an axial-azimuthal section of a SPT-100 Hall thruster. $x$ is along the axial direction and $y$ is oriented along the azimuth. The $z$-coordinate represents the radial direction, which is not self-consistently resolved in any of the simulations in this paper. The domain is 5 cm-long in the $x$-direction and 1 cm-long in the $y$-direction. The thruster's discharge channel extends from $x = 0$ to $x = 2.5$ cm.

For the 1D axial simulations, only the $x$-coordinate is resolved. Therefore, to incorporate the effect of azimuthal instabilities in enhancing the cross-field transport of the electrons, a Bohm-type transport model of the general form $v_{Bohm} = \alpha \omega_{ce}$, similar to that used in the reference LANDMARK case [25], was adopted. In the relation for the transport model, $v_{Bohm}$ is the Bohm Collision frequency, $\omega_{ce}$ is the electron cyclotron frequency, and $\alpha$ is a tuning parameter. The values of $\alpha$ for inside the discharge channel and for the near-plume are set as suggested in the LANDMARK case, i.e., 0.1 and 1.0, respectively. This selection of the $\alpha$ values yields the Bohm collision frequency profile shown in Figure 1(a).

Figure 1(a) also presents the axial distribution of the magnetic field intensity ($B$), which is used for the 1D axial and Q2D axial-azimuthal simulations. Consistent with the LANDMARK case, the magnetic field has a bi-Gaussian profile, and its maximum intensity is 15 mT, occurring at the channel's exit plane ($x = 2.5$ cm) [25].

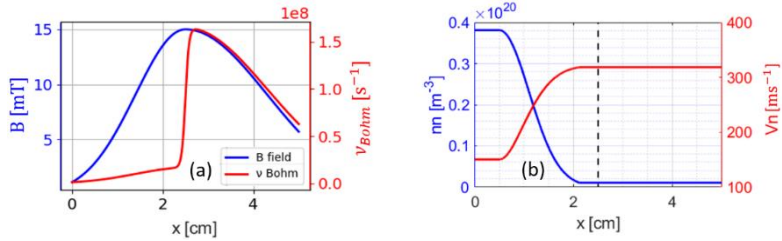

Figure 1: Axial profiles of several simulations' parameters: (a) the magnetic field intensity ($B$) and the Bohm collision frequency ($v_{Bohm}$) for the 1D axial simulations, (b) the initial number density and velocity of the neutrals

Initially, the ions and electrons are loaded uniformly throughout the domain at a density of $5 \times 10^{17}$ $m^{-3}$. The particles are sampled from a Maxwellian at a temperature of 0.5 eV for the ions and 10 eV for the electrons. The neutrals are loaded according to the number density and velocity profiles shown in plot (b) of Figure 1. For the simulations where the neutrals are modeled kinetically, they are sampled from a Maxwellian at an initial load temperature of 640 K.

Regarding the axial boundary conditions for the plasma potential, a Dirichlet boundary condition of 300 V is applied at the anode side of the domain ($x = 0$), whereas at the cathode side ($x = 5$ cm), the potential is set at 0 V. For the Q2D simulations, a periodic boundary condition is applied at the azimuthal ends of the domain. The potential is solved using the Thomas Tridiagonal algorithm in the 1D axial simulations, whereas the solution of the plasma potential in the Q2D simulations is obtained from the Reduced-Dimension Poisson Solver (RDPS) [23] that solves a reduced-order Poisson system of equations using a direct matrix solve algorithm based on the LU decomposition.

The particles' boundary conditions are implemented as follows: along the axial direction, electrons leaving the domain at the anode are removed from the simulation. Except for the simulations in Section 4.3, the ion recombination is considered in all cases for the ions reaching the anode. For the 1D axial simulations with kinetic-DSMC treatment of the neutrals, the neutral particles reaching the anode surface are isotropically reflected into the domain. In addition, the recombined neutrals are sampled from a Maxwellian at their initial load temperature. In the simulation cases with neutrals' fluid description, however, the flux of recombined neutrals is added to the anode particle flow rate. The ions leaving the domain from the cathode side are removed from the simulation.

To sustain the discharge, electrons are injected from the cathode boundary into the domain according to the quasi-neutrality condition at the cathode [15]. The injected electrons are sampled from a half-Maxwellian at the initial electrons' load temperature, i.e., 10 eV.

For the Q2D simulations, a periodic particles' boundary condition is used; hence, particles exiting the domain along the azimuthal direction from one side are re-introduced into the domain from the opposite side.

Concerning the interactions among the particles, we have taken into account the electron-neutral and ion-neutral collisions. The neutral-neutral interactions are also explicitly modeled as Variable Hard Sphere collisions in the simulations where the neutral species is treated using the kinetic-DSMC approach (Section 4.2).

The collisions between the electrons and neutrals are resolved following the Monte Carlo Collisions (MCC) algorithm [26]. In this regard, to be consistent with the reference LANDMARK case, only the ionization and the elastic momentum exchange collision are considered with the corresponding cross-sections reported in Ref. [25]. In the case of each collision event, the electrons are isotopically reflected. Moreover, when an ionization event occurs, the energy loss is also considered for the incident electron following the approach explained in Ref. [25].

For the ion-neutral collisions, we have considered the elastic momentum and charge exchange collisions. When the neutrals are modeled as a fluid, the MCC algorithm is used to resolve the collisions between the ions and the neutrals. However, when the neutrals are treated as particles, the ion-neutral collisions are resolved following the DSMC algorithm. More details about the handling of the ion-neutral collisions in the simulations carried out, including the collision cross-sections used, are provided in Section 3.2.

| | Value [unit] | |
|---|---|---|
| **Parameter** | **1D axial** | **Q2D axial-azimuthal** |
| **Computational parameters** | | |
| Domain's axial length ($L_x$) | 5 [cm] | 5 [cm] |
| Domain's azimuthal length ($L_y$) | --- | 1 [cm] |
| Domain's real cross-sectional area ($A_{real}$) | 40.035 [$cm^2$] | 40.035 [$cm^2$] |
| Axial cell size ($\Delta x$) | 20 [μm] | 16.6 [$\mu m$] |
| Azimuthal cell size ($\Delta y$) | --- | 16.6 [$\mu m$] |
| Number of cells in axial direction ($N_i$) | 2500 | 3000 |
| Number of cells in azimuthal direction ($N_j$) | --- | 600 |
| Time step ($ts$) | $2.5 \times 10^{-12}$ [$s$] | $2 \times 10^{-12}$ [$s$] |
| Initial number of macroparticles per cells for the axial grid ($N_{ppc}$) | 50 | 50 |
| Total simulated time | 250 [$\mu s$] | 200 [$\mu s$] |
| **Physical parameters** | | |
| Initial plasma number density ($n_{i,0}$) | $5 \times 10^{17}$ [m$^{-3}$] | $5 \times 10^{17}$ [m$^{-3}$] |
| Initial electron temperature ($T_{e,0}$) | 10 [$eV$] | 10 [$eV$] |
| Initial ion temperature ($T_{i,0}$) | 0.5 [$eV$] | 0.5 [$eV$] |
| Neutral temperature ($T_n$) | 640 [$K$] | 640 [$K$] |
| Anode mass flow rate ($AMFR$) | 5 [mgs$^{-1}$] | 5 [mgs$^{-1}$] |
| Anode voltage ($V_a$) | 300 [$V$] | 300 [$V$] |
| Ad-hoc wall collision frequency ($\nu_{wall}$) | $10^7$ [$s^{-1}$] | $5 \times 10^6$ [$s^{-1}$] |

Table 1: Summary of the computational and physical parameters for the 1D axial and Q2D axial-azimuthal simulations

Finally, apart from the simulations whose results are reported in Section 4.4, all simulations resolve the electron-wall collisions and the associated energy losses in an empirical manner, following the approach explained in Ref. [27] and also adopted in the LANDMARK simulation case. The electron-wall collisions are implemented as a fictitious collision event in the MCC module with the ad-hoc frequency of $\nu_{wall} = 1 \times 10^7\ s^{-1}$ for the 1D axial simulations and $v_{wall} = 5 \times 10^6 \ s^{-1}$ for the Q2D simulations. The reason why the wall collision frequency for the axial-azimuthal simulations is half of that for the 1D axial simulations is explained in Section 4.5. If an electron-wall collision event occurs, the particle is isotropically reflected and loses some of its energy, which is calculated from the relation

$$\mathcal{W} = \ \epsilon \exp\left(-\frac{\mathcal{U}}{\epsilon}\right), \tag{Eq. 1}$$

in which $\epsilon$ is the electron's kinetic energy in eV, and $\mathcal{U}$ is a threshold electron energy that is set at 20 eV to be consistent with the LANDMARK case description [25].

Following the above descriptions of the simulations' setup, Table 1 summarizes the main computational and physical parameters that we have used for the simulations in this work. For reference, it is pointed out that, based on our choices of the physical parameters, the performed simulations correspond to Case 1 among the LANDMARK's three simulation cases for the fluid/hybrid codes (Test Case 3) [25].

## Section 3: Description of the fluid and kinetic models for the neutrals

In this section, we describe in detail the models that we used to resolve the neutral dynamics in the simulations. We present, in the next section, the 1D fluid system of equations for the neutrals and, in Section 3.2, we discuss the implementation of the DSMC module that was used to model the neutral-neutral and the ion-neutral collisions in the simulations with kinetic-DSMC treatment of the neutral particles.

### 3.1. Neutrals' fluid model

The neutrals in a Hall thruster can be approximated as an adiabatic inviscid flow. Hence, assuming that the variation in the neutrals' properties is mostly along the axial direction, their spatiotemporal evolution can be described through a 1D Euler system of equations in the following form

$$\frac{\partial \rho}{\partial t} + \frac{\partial (\rho u)}{\partial x} = S_1, \tag{Eq. 2}$$

$$\frac{\partial (\rho u)}{\partial t} + \frac{\partial (\rho u^2)}{\partial x} = -\frac{\partial P}{\partial x} + S_2, \tag{Eq. 3}$$

$$\frac{\partial \rho E}{\partial t} + \frac{\partial (\rho u E)}{\partial x} = -\frac{\partial P u}{\partial x} + S_3. \tag{Eq. 4}$$

In the above equations, $\rho$ is the neutral density, $u$ is the axial drift velocity, $P$ is the pressure, and $E = e + \frac{1}{2}u^2$ is the specific total energy with $e$ being the specific internal energy, $e = \frac{n k_B T}{\rho(\gamma-1)}$. In the relation for $e$, $n$ is the number density, $k_B$ is the Boltzmann constant, $T$ is the neutral temperature, and $\gamma$ is the heat capacity ratio. Using the ideal gas relation, $P = n k_B T$, the specific internal energy can be expressed as $e = \frac{P}{\rho(\gamma-1)}$.

In Eqs. 2 to 4, $S_1$, $S_2$, and $S_3$ are the source terms for the continuity, momentum, and energy equations, respectively, which we will define shortly. To numerically solve the system of equations given by Eqs. 2-4, we write them in a general conservation form as

$$\frac{\partial U}{\partial t} + \frac{\partial \Gamma(U)}{\partial x} = S(U), \tag{Eq. 5}$$

with the discretized form

$$U_i^{t+1} = U_i^t - \frac{\Delta t}{\Delta x}\left(\Gamma_{i+\frac{1}{2}} - \Gamma_{i-\frac{1}{2}}\right) + S \Delta t. \tag{Eq. 6}$$

In Eq. 6, $U$ is the state vector of the conserved quantities, $\Gamma(U)$ is the vector of the corresponding fluxes, and $S(U)$ is the vector of the source terms. These vectors are defined as

$$U = \begin{pmatrix} \rho \\ \rho u \\ \rho E \end{pmatrix}, \quad \Gamma = \begin{pmatrix} \rho u \\ \rho u^2 + P \\ \rho u E + P u \end{pmatrix}, \quad S = \begin{pmatrix} -S_i \\ -u S_i \\ -u^2 S_i \end{pmatrix}. \tag{Eq. 7}$$

In the discretized form of the equations (Eq. 6), $U_i^t$ is the state vector value at time $t$, $\Gamma_{i-\frac{1}{2}}$ and $\Gamma_{i+\frac{1}{2}}$ are the flux vectors at the left and right boundaries of the $i$-th cell, $\Delta t$ is the timestep and $\Delta x$ is the cell size. Moreover, $S_i$ in Eq. 7 is the ionization source term defined as $\rho \nu_{ion}$, where $\nu_{ion}$ is the ionization frequency. It is noted that the energy change in the neutral flow due to the ionization is assumed to be negligible and is, thus, neglected for the simulations in this work.

The described 1D Euler system of equations can be simplified by assuming the neutrals as an isothermal flow, which is a typical assumption made in Hall thrusters' modeling [6][15]. This assumption implies that the energy equation is no longer required in the Euler system and can be replaced with the equation of state, $P = \frac{\rho}{M} k_B T$, in which $M$ is the atomic mass.

From Eq. 6, it is evident that a first-order Euler scheme is chosen for the time marching. Nevertheless, as part of the benchmarking of the neutrals' fluid model against the Sod shock tube problem (See Section A.1 of the Appendix), we also carried out the simulations with the 3rd-order Runge Kutta scheme for the time advancement and obtained the same results as that with the first-order Euler scheme. To solve the system spatially, we adopted the Weighted Essentially Non-Oscillatory scheme with 5 grid points (i.e., WENO-5). To calculate the inter-cell fluxes, we used the HLLC Riemann solver [28]. The algorithms used for these schemes are implemented as described in Ref. [29]. The neutrals' fluid model is also verified in a simplified Hall thruster problem as described in Section A.2 of the Appendix.

The initial profiles for the neutral density and velocity are shown in Figure 1(b). Moreover, a uniform neutral temperature ($T_n$) of 640 K is assumed. The neutrals' fluid model in this work is purely 1D and axial with no attempt made to include the effect of neutral-wall collisions into the model in an ad-hoc manner similar to the approach pursued in Ref. [30]. Investigation of the effect(s) of adding such term to the model is postponed to the future work.

Concerning the boundary conditions, at the cathode side of the domain, it is reasonable to apply a Neumann boundary condition as in Eq. 8 because the neutral flow is supersonic at the cathode boundary.

$$\frac{\partial U}{\partial x}\bigg|_{x=L_x} = \frac{U_{N_i} - U_{N_i-1}}{\Delta x} = 0 \implies U_{N_i} = U_{N_i-1}. \tag{Eq. 8}$$

At the anode boundary of the domain, a constant mass flow rate ($\dot{m}_a$) of neutrals is injected into the domain together with an additional variable mass flow rate ($\dot{m}_r$) that corresponds to the flux of ions recombined on the anode surface. Hence, following the approach suggested in Ref. [15], the flow properties at the anode boundary (i.e., the first computation cell) at the time $t + 1$ can be written as

$$u_1^{t+1} = u_1^t + \Delta u \,, \tag{Eq. 9}$$

$$\rho_1^{t+1} = \frac{\dot{m}_a + \dot{m}_a^t}{u_1^{t+1} A} \,, \tag{Eq. 10}$$

$$P_1^{t+1} = \frac{\rho_1^{t+1}}{M} k_B T_{n0} \,. \tag{Eq. 11}$$

In Eq. 9, $\Delta u$ is defined as in Eq. 12 using the flow properties in the first (subscript 1) and the second (subscript 2) cells of the domain

$$\Delta u = \frac{u_1^2 (u_1 - C_{s1})[P_2 - P_1 - \rho_1 C_{s1}(u_2 - u_1)]}{\left(\frac{\dot{m}}{A} K_B N_A T_{n0} - \rho_1 C_{s1}\right)\left(\frac{\Delta x}{\Delta t} - (u_1 - C_{s1})\right)}. \tag{Eq. 12}$$

In above relations, $A$ is the cross-sectional area, $N_A$ is the Avogadro's number, and $C_{s1}$ is the sound speed in cell 1, defined as $\sqrt{\frac{\gamma P_1}{\rho_1}}$ with $\gamma$ being the heat capacity ratio.

In this work, we have performed simulations with several variants of the neutrals' fluid model described above. In the simplest case, we assume that the velocity profile of the neutrals remains invariant throughout the simulation and, thus, we only solve the continuity equation to obtain the temporal variation in the neutral density as a result of their depletion due to the ionization and their replenishment due to the gas injection and ion recombination on the anode surface. This model is referred to as the "Continuity only" in this paper.

Increasing the complexity level by one step, the second neutrals' fluid model that we used is the simplified version of the 1D Euler system, which accounts for the time variation of the density and velocity profiles of isothermal neutrals. We call this model the "1D Euler" in the rest of the paper.

Finally, the most complex fluid description of the neutrals considered in this work solves the full 1D Euler system, formulated in Eqs. 2-4, and additionally takes into account the axial momentum transferred from the ions to the neutrals through the ion-neutral collisions. This model is referred to as the "full 1D Euler + ion momentum transfer" in the text. The reason why the full Euler system is needed when considering the ion-neutral momentum transfer will be clarified in Section 4.1 where we have also presented the simulations' results with the isothermal continuity-plus-momentum neutrals' fluid model in the presence of the ion-neutral collisions, referred to as the "1D Euler + ion momentum transfer". It is noted that the "1D Euler + ion momentum transfer" model for the neutrals had been suggested recently in the literature [15]. The effect of the ion-neutral collisions on the momentum of the neutral atoms is introduced in the model as another source term ($S_m$) added to the right-hand side of Eq. 3. The energy transfer associated with ion-neutral collisions is also included as a source term in Eq. 4 in the form of $uS_m$.

The ion-neutral collisions, where considered, are simulated through the MCC method. The ion-neutral momentum exchange (MEX) and charge exchange (CEX) collisions are considered. The cross-sections for these collisions are plotted in Figure 2. The momentum source term $S_m$ can be either approximated as $-\rho_i(v_i - u)\nu_{in}$, where $v_i$ is the ions' axial drift velocity and $\nu_{in}$ is the total ion-neutral collision frequency, or to be explicitly computed as the difference between the total axial momentum of the ions' population prior and after the collisions. The latter approach is pursued in this work.

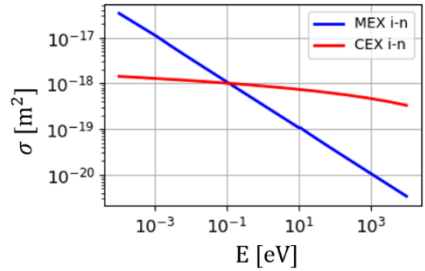

Figure 2: Ion-neutral collision cross sections used in the simulations that resolve the ion-neutral interactions. Data from Ref. [25].

As a final remark, the full 1D Euler system given by Eqs. 2-4 was found to be less numerically robust compared to the other neutrals' fluid models mentioned above that do not include the energy equation. In this respect, to obtain stable solutions from the full Euler system, more stringent CFL (Courant–Friedrichs–Lewy) criterion was used. In addition, the system was very sensitive to the noise in the $S_m$ source term in the energy equation. Thus, for the 1D axial and Q2D axial-azimuthal simulations with the "full 1D Euler + ion momentum transfer" neutrals' fluid model, a moving-average filter with the window size of 100 was applied to the $S_m$ source term before passing the corresponding array to the neutrals' dynamics function.

### 3.2. Neutrals' kinetic model with the DSMC collisions

In the simulations where the neutral dynamics was modeled kinetically, i.e., in the same way that the plasma species were treated, a Direct-Simulation Monte-Carlo (DSMC) module is implemented to simulate the collisions involving heavy particles. In this module, three types of collisions are considered that include the elastic collisions between the neutrals (n-n) as well as the isotropic momentum exchange (MEX i-n) and the charge exchange (CEX i-n) collisions between the ion and neutral particles. The cross-section for the momentum exchange collisions between the neutrals is obtained according to Bird's Variable Hard Sphere (VHS) model [31]

$$\sigma = \pi \left( d_{ref} \left( \frac{v_{r,ref}}{v_r} \right)^\gamma \right)^2, \tag{Eq. 13}$$

in which the values of $d_{ref}$, $v_{r,ref}$, and $\gamma$ are, respectively, $5.74 \times 10^{-10}$ m, 209.8 m/s, and $-0.7$. $v_r$ is the relative velocity between the colliding particle pair. The cross-sections for the MEX and CEX collisions are the ones plotted in Figure 2.

The DSMC algorithm starts with the sorting of the particles based on the cells in which they reside. In each cell, the maximum number of collision events (or collision pairs ($n_p$)) for each collision type is calculated as per the "no-time-counter" method (NTC) [32]

$$n_p = N_1\, N_2\, W(v_r\sigma)_{max}\frac{\Delta t}{V_{cell}}.$$  (Eq. 14)

In the above equation, $N_1$ and $N_2$ are the number of macroparticles of each species within the cell, $V_{cell}$ is the cell volume, and $W$ is the macroparticle weight of the species with the larger macroparticle weight, which, for the simulations in this paper, corresponds to the neutrals' population. $(v_r\sigma)_{max}$ is initialized with an estimation and gets updated during the simulation when a larger value than the estimated one is encountered. The number of collision pairs from the expression in Eq. 14 must be halved if the collision happens between two particles of the same species. Therefore, the total number of collision pairs in each cell is obtained as

$$n_p = \left(\frac{1}{2}N_n^2\, W_n(v_r\sigma)_{max,n} + N_nN_i\, W_n(v_r\sigma)_{max,MEX} + N_nN_i\, W_n(v_r\sigma)_{max,CEX}\right)\frac{\Delta t}{V_{cell}},$$  (Eq. 15)

where the indices "$n$" and "$i$" represent the neutrals and the ion species, respectively. To pick a collision pair in each cell, two numbers $p_1 \in [1, N_n]$ and $p_2 \in [1, N_n + 2N_i]$ are randomly selected, where $N_n$ and $N_i$ are, respectively, the number of neutral and ion macroparticles in that cell. The type of collision for this pair is determined using a Monte Carlo method as illustrated in Figure 3. The collision probability for each pair is defined as

$$P_i = \frac{v_r\sigma}{(v_r\sigma)_{max}},$$  (Eq. 16)

where $\sigma$ is the cross-section of the selected collision type. To check for collision occurrence, the probability $P_i$ from Eq. 16 is compared with a uniformly distributed random number.

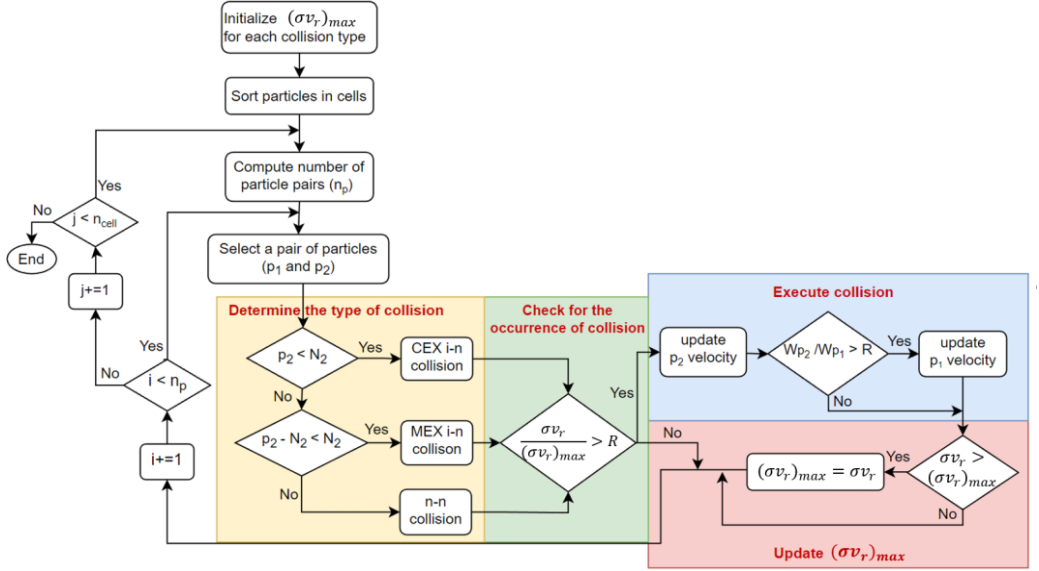

Figure 3: A Flowchart illustrating the implementation of the DSMC collisions algorithm.

In case of the neutral-neutral or neutral-ion MEX collisions, the post-impact velocities of the involved particles are obtained based on the conservation of momentum and energy

$$\vec{v}'_{r1} = \frac{m_1\vec{v}_1 + m_2\vec{v}_2}{m_1 + m_2} + \frac{m_2}{m_1 + m_2}\, v_r\vec{R},$$  (Eq. 17)

$$\vec{v}'_{r2} = \frac{m_1\vec{v}_1 + m_2\vec{v}_2}{m_1 + m_2} - \frac{m_1}{m_1 + m_2}\, v_r\vec{R},$$  (Eq. 18)

where $\vec{R}$ is a unit vector of random direction. Assuming hard sphere interaction between the particles, the scattering will be isotropic. Therefore, the scattering angles ($\psi$ and $\phi$) are defined as

$$\psi = \cos^{-1}(2\mathcal{R}_1 - 1)\ ,\quad \phi = 2\pi\mathcal{R}_2\ ,\ \text{ where }\ \mathcal{R}_1 \text{ and } \mathcal{R}_2 \in [0,1)\ .$$  (Eq. 19)

In case of the neutral-ion CEX collision, the velocities of the two particles are simply swapped to update the colliding particles' velocity.

Eqs. 17 and 18 are only true when the colliding particles are of equal macroparticle weights. However, to ensure the conservation of momentum and energy in case of unequal macroparticles weights over statistically large number of collision events, the velocity update is always performed for the particle with the smaller macroparticle weight ($W_{p_2}$), whereas the velocity of the particle with the larger macroparticle weight ($W_{p_1}$) is modified only if $\mathcal{R} > W_{p_2}/W_{p_1}$, where $\mathcal{R}$ is a random number.

In the simulations of this effort, the neutrals' macroparticle weight is ten times that of the charged particles. As also pointed out in Section 2, the neutral particles are initially loaded in the domain following the density and drift velocity profiles in Figure 1(b). Throughout a simulation, the neutrals are injected with a constant anode mass flow rate ($\dot{m}_a$) at the anode surface with an assumed axial drift velocity of $150\ m/s$. Since the radial coordinate is not resolved, no neutral-wall interaction is accounted for.

Two versions of the described DSMC collisions model are compared in the 1D axial simulations. In the first version, we only consider the neutral-neutral collisions, whereas, in the second version, the ion-neutral MEX and CEX collisions are also taken into account.

Finally, as it will be seen in Section 4, to highlight the effect of neutrals' collision on the dynamics of the plasma discharge, we have also performed, for reference, additional simulations with kinetic neutrals' treatment in the absence of any neutral collisions and with various neutral injection drift velocities.

## Section 4: Results and discussion

Results of the 1D axial and Q2D axial-azimuthal simulations carried out with various models for the neutral dynamics are discussed in this section. We start by presenting the 1D axial PIC simulation results with the fluid and kinetic-DSMC neutrals' models in Sections 4.1 and 4.2. In Sections 4.3 and 4.4, we evaluate, in the 1D axial configuration, the impact of the ion recombination at the anode surface and the electron-wall collisions on the dynamics of the discharge when using different fluid models of the neutrals. In Section 4.5, the results from the Q2D axial-azimuthal simulations are presented to assess the influence of the neutrals' model on the plasma dynamics in simulations with self-consistent resolution of the electrons' cross-field mobility. Finally, we compare the global performance parameters obtained from various simulations in Section 4.6.

### 4.1. 1D axial PIC simulations with different neutrals' fluid models

Figure 4 and Figure 5 show the spatiotemporal maps of the ion number density and the ionization rate, respectively, from the 1D axial simulations with different fluid models for the neutrals.

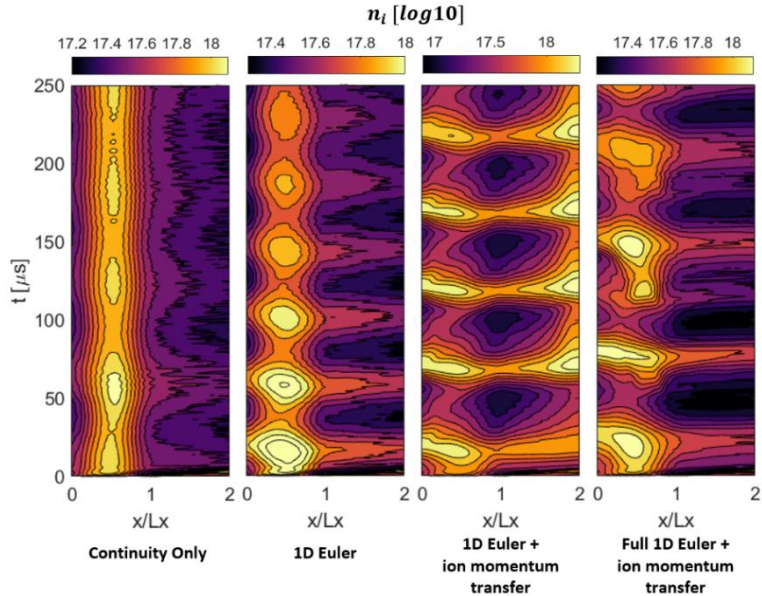

Figure 4: Spatiotemporal evolutions of the ion number density from 1D axial simulations with different neutrals' fluid models

Similarly, Figure 6 and Figure 7 presents the variations in time and space of the neutral flow properties, namely, the number density and velocity. From these four figures, it is noticed that, in case of the "Continuity Only" model, the breathing oscillations are rather weak and are damped early in the simulation. The discharge in this case becomes almost steady from approximately $150 \ \mu s$.

In contrast, we observe from the spatiotemporal evolution plots in Figure 4 and Figure 5 that resolving the variations in the neutral velocity in the "1D Euler" model results in more distinguishable breathing oscillations. Additionally, taking into account the ion-neutral momentum transfer in the "1D Euler + ion momentum transfer" case yields stronger breathing oscillations that persist for a longer time compared to the "1D Euler" simulation. In any case, the discharge oscillations show an overall damping behavior in these two cases as well.

The pattern of the oscillations from the simulation with "Full 1D Euler + ion momentum transfer", where the isothermality assumption is not made for the neutral flow, is seen to be relatively different compared to that from the simulations with "1D Euler" and "1D Euler + ion momentum transfer" neutrals' models. Additionally, the frequency of the oscillations from the "Full 1D Euler + ion momentum transfer" simulation is noticed to be lower than the simulation with "1D Euler + ion momentum transfer" model.

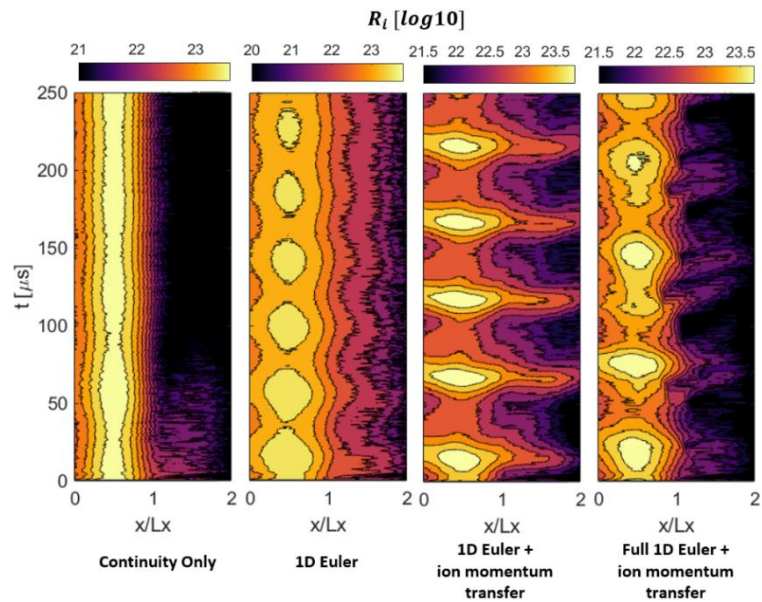

Figure 5: Spatiotemporal evolutions of the ionization rate from 1D axial simulations with different neutrals' fluid models

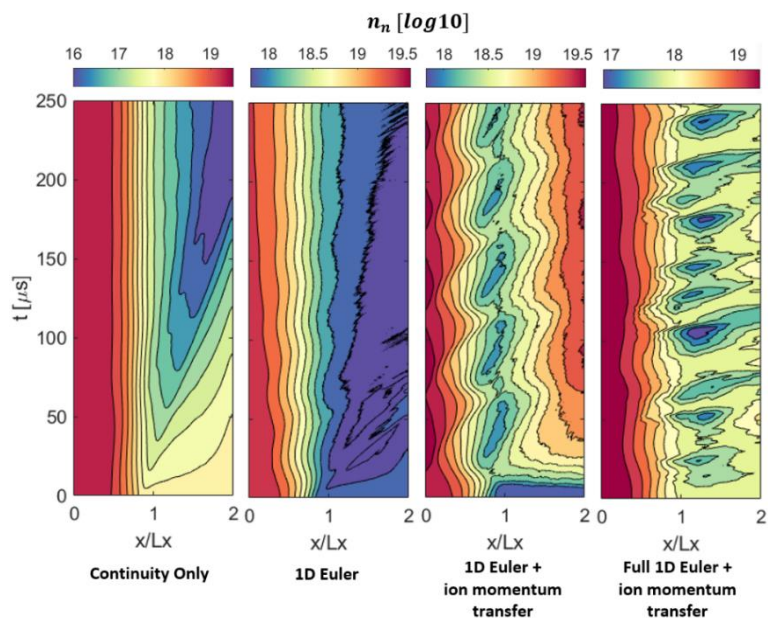

Figure 6: Spatiotemporal maps of neutral number density from 1D axial simulations with different neutrals' fluid models

Referring to Figure 4 to Figure 7, it is also interesting to note that, in case of the "1D Euler + ion momentum transfer" and "full 1D Euler + ion momentum transfer" models, the oscillatory behavior in the spatiotemporal maps of the ion number density and the ionization rate are reflected in the neutral number density and velocity evolutions as well. This is a consequence of the ion-neutral collisions and the resulting momentum transfer that can further couple the plasma and neutral dynamics.

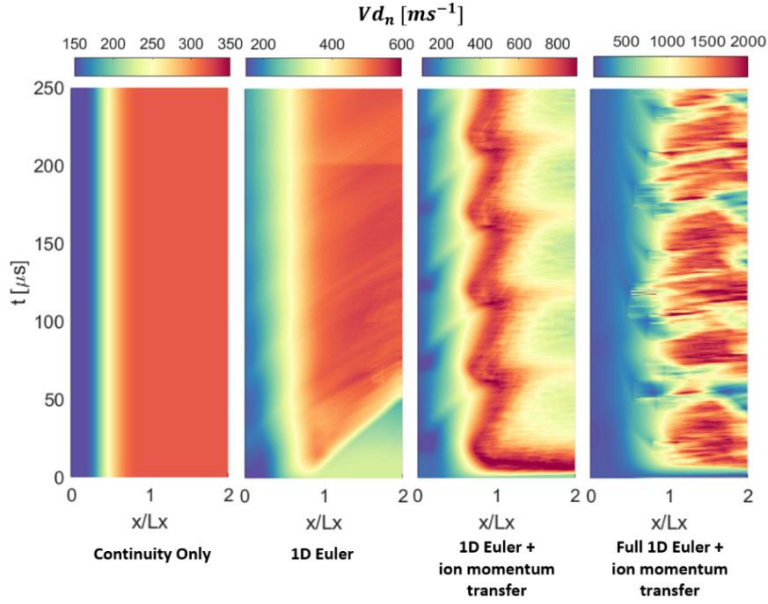

Figure 7: Spatiotemporal maps of neutral axial velocity from 1D axial simulations with different neutrals' fluid models

Another noteworthy point concerning Figure 4 and Figure 6 is the notable increase in the ion and neutral number density is the plume in the simulation case with the "1D Euler + ion momentum transfer" neutrals' model. The increased neutral density in the plume has led to the extended spatial distribution of the ionization rate seen to periodically occur in Figure 5. Furthermore, an elevated neutral density in the plume unrealistically enhances the ion-neutral collisions. The combined effect of the increased ionization and exaggerated ion-neutral momentum transfer in the plume has translated into the quite high ion densities in this zone. These unrealistic results from the "1D Euler + ion momentum transfer" simulation can be linked to the isothermality assumption for the neutrals since it is observed that the simulation with the "full 1D Euler + ion momentum transfer" neutrals' fluid model does not yield the same predictions.

In Figure 8, the time-averaged axial distributions of several plasma and neutral properties are shown from the 1D axial simulations with different neutrals' fluid models. The available reference profiles from LANDMARK's 1D fluid simulation are superimposed as dashed black curves on the plots in the first and second column of Figure 8.

First, it is seen that the results of our 1D PIC simulations with "Continuity Only" model (the blue curves) are in great agreement with the reference profiles. The reason behind the similarity of results between these two cases is that, also in the LANDMARK simulation, only the evolution of the neutrals' density was resolved. The only major difference is noticed to be in the axial electric field profile (Figure 8(a)) where the discontinuity in the LANDAMRK's profile is absent in our result. This is associated with the kinetic treatment of the electrons in our simulations compared to the fluid description of the electron species in the LANDMARK [25].

Second, looking at the profiles from the simulation with "1D Euler" neutrals' model (solid red curves), we observe that the electric field and plasma potential distributions (Figure 8(a) and (g)) do not show significant variations compared to the corresponding profiles from the "Continuity Only" case. Nevertheless, as the neutrals' velocity (Figure 8(i)) in the "1D Euler" case is overall larger compared to the "Continuity Only" case, the peak of the ionization rate (Figure 8(e)) and, consequently, the peak ion number density (Figure 8(b)) as well as the neutral number density (Figure 8(h)) near the anode are all lower in the "1D Euler" case. Nevertheless, since the ionization rate is relatively higher around the exit plane (vertical dashed black line) and into the near-plume in the "1D Euler" case, the electron kinetic energy is lower in those regions with respect to the "Continuity Only" case.

Third, the time-averaged profiles from the "full 1D Euler + ion momentum transfer" simulation case, illustrated as solid green curves in Figure 8, are in most cases quite similar to the profiles from the "1D Euler" simulation.

Particularly, the profiles of the electric field, plasma potential, and electron energy, which are mostly related to the dynamics of the electron species, are seen to be rather unaffected by the ion-neutral collisions.

The main difference between the profiles of the simulations with the "1D Euler" and "full 1D Euler + ion momentum transfer" neutrals' models is expectedly observed in the plume where the ion-neutral collisions are noticed to have resulted in a significant increase in the neutral velocity (Figure 8(i)), an overall increase in the ion temperature (Figure 8(f)), and a slight decrease in the ion axial velocity (Figure 8(c)). The interactions between the ions and the neutrals near the anode have led to an increase in the neutral density in this zone (Figure 8(h)), resulting in a slightly enhanced ionization (Figure 8(e)) and, thus, an increased ion (plasma) number density in the near-anode zone (Figure 8(b)).

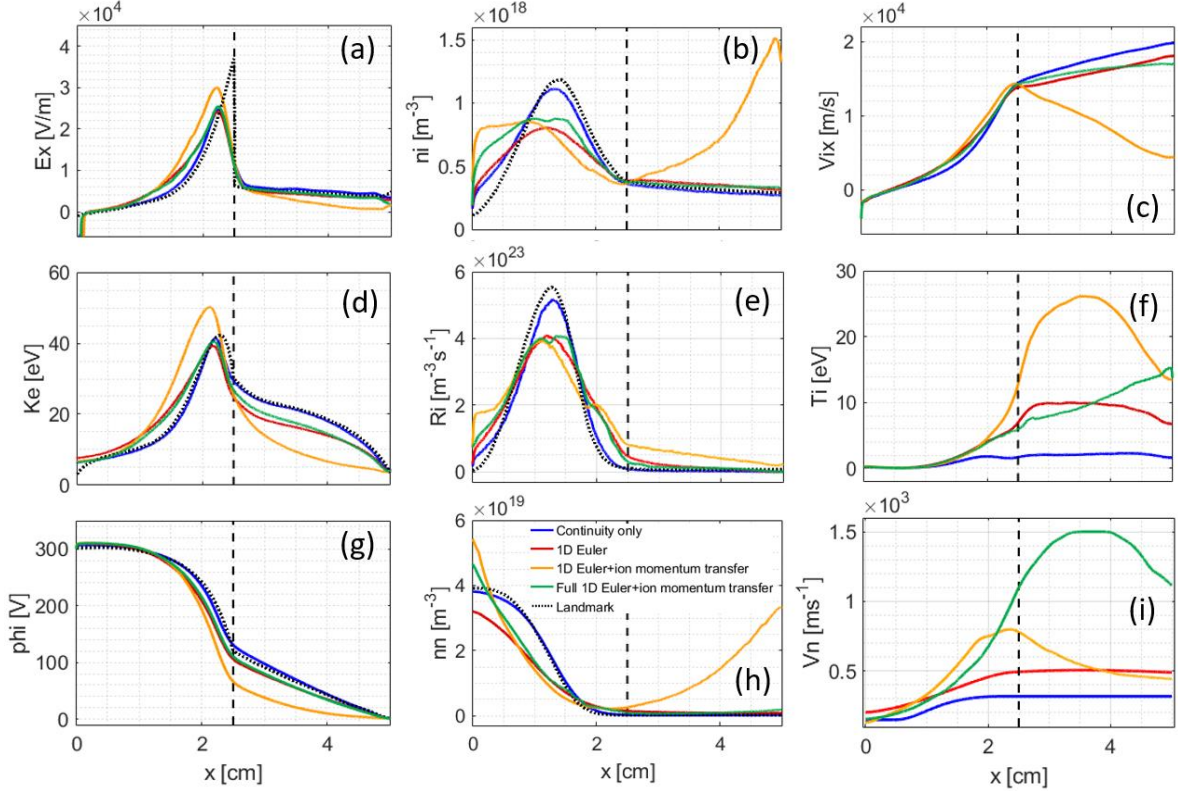

Figure 8: Time-averaged (over 250 $\mu s$) axial profiles of the plasma and neutral properties from 1D axial PIC simulations with different neutrals' fluid models. The LANDMARK results from Ref. [25] are superimposed as dashed green curves. (a) axial electric field ($E_x$), (b) ion number density ($n_i$), (c) ion axial velocity ($V_{ix}$), (d) electron kinetic energy ($Ke$), (e) ionization rate ($R_i$), (f) ion temperature ($T_i$), (g) electric potential ($phi$), (h) neutral number density ($n_n$), and (i) neutral axial velocity ($V_n$).

Fourth, the profiles from the simulation with isothermal "1D Euler + ion momentum transfer" neutrals' model (solid yellow curves) show notable differences with respect to the profiles obtained using the other neutrals' models. In this regard, it is seen in plot (c) of Figure 8 that the ion axial velocity is reduced significantly in the plume. Moreover, the ion number density (Figure 8(b)) is largely increased toward the cathode, which was also observed from the spatiotemporal evolution of the ion density for the case of this neutrals' fluid model (Figure 4). In addition, the neutral density (Figure 8(h)) shows a significant rise in the plume, whereas the neutral velocity (Figure 8(i)) has dropped after the exit plane.

The observed variations in the $n_i$ and the $V_{ix}$ in the plume due to the ion-neutral collisions are obviously exaggerated in the "1D Euler + ion momentum transfer" simulation because of the large neutral number density in the plume. In this regard, the main reason behind this significant neutral density and a surprising decrease in the neutral velocity in the plume, despite the neutrals gaining momentum from the accelerated ions, is hypothesized to be the occurrence of an unexpected shock in the neutral flow in the isothermal "1D Euler + ion momentum transfer" simulation case. We revisit this hypothesis in Section 4.2.1. In any case, a shock wave decelerates the neutral flow while increasing its pressure. The increase in pressure in the "1D Euler + ion momentum transfer" simulation case directly translates into a rise in the number density because of our isothermality assumption for the neutral flow (Section 3.1). In this respect, the abrupt changes expected in the neutral properties due to a shock are not visible in the relevant plots in Figure 8 because the ionization of the neutrals smooths out the potential discontinuities over time.

Figure 9 presents the time evolution plots of the discharge current ($I_d$), the ion current ($I_i$), and the electron current ($I_e$) from the 1D axial PIC simulations with the "Continuity Only" (plot (a)), the "1D Euler" (plot (b)), the "1D Euler + ion momentum transfer" (plot (c)), and the "full 1D Euler + ion momentum transfer" neutrals' models. The various current terms are calculated based on the approach presented in Ref. [22]. Referring to the plots in Figure 9, it is evident that the amplitude of the current oscillations consistently increases from the "Continuity Only" simulation in plot (a) to the "1D Euler + ion momentum transfer" one in plot (c). The overall evolution of the current terms from the "full 1D Euler + ion momentum transfer" simulation (plot(d)) is, nonetheless, quite different compared to the other cases. Indeed, the amplitude of the current oscillations is more comparable with the "1D Euler" simulation whereas the damping of the oscillations occurs over a longer timescale compared to the "1D Euler" case but slightly shorter than that for the "1D Euler + ion momentum transfer" simulation.

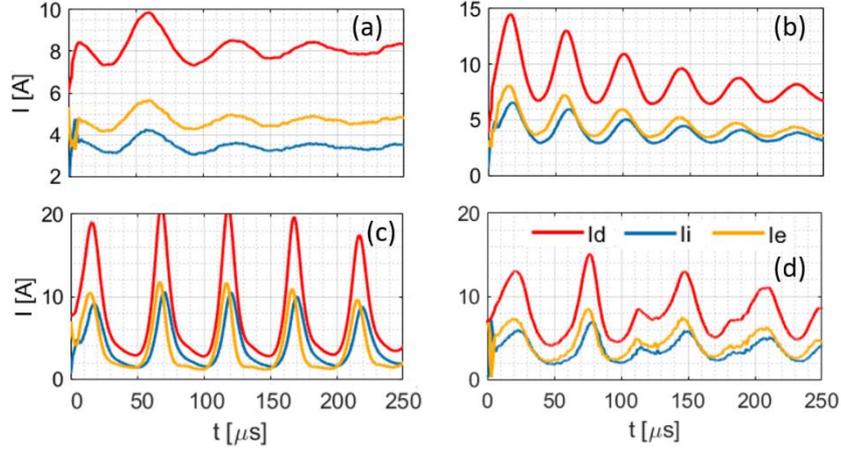

Figure 9: Time evolution of various current terms from the 1D axial simulations with different neutrals' fluid models; (a) "Continuity Only", (b) "1D Euler", (c) "1D Euler + ion momentum transfer", (d) "full 1D Euler + ion momentum transfer".

### 4.2. 1D axial PIC simulations with neutrals' kinetic treatment

In this section, we present the results from 1D axial simulations in which the neutrals are treated as particles and their evolution is kinetically resolved. The following results allow us to evaluate the extent to which the simulated dynamics of the discharge with the fluid descriptions of the neutrals is representative of the resolved dynamics when using higher fidelity kinetic neutrals' model.

#### 4.2.1. Results of the simulations with the DSMC collisions

Figure 10 shows the spatiotemporal evolutions of the ion number density, neutral number density, and neutral velocity over 230 $\mu s$ of simulated time from the 1D axial simulations with the DSMC collisions.

The main point to highlight in Figure 10 is that, in case of both the neutral-neutral collisions ("n-n DSMC") and neutral-neutral plus ion-neutral collisions ("n-n + i-n DSMC"), the discharge is seen to reach a steady state with no to weak global oscillations in the evolution maps of the discharge properties. More specifically, it is observed that considering only n-n collisions results in the oscillatory patterns to persist for a longer time whereas, in the presence of ion-neutral collisions as well, the discharge becomes steady after a high-amplitude initial transient.

The discharge behavior described above for the simulations with different DSMC collisions is also reflected in the time evolution plots of various current terms as shown in Figure 11. In addition, an interesting point from plot (b) in Figure 11 is that the mean value of the current terms from the axial simulation with the n-n collisions only is very similar to the corresponding values from the axial simulation with the "1D Euler" neutrals' fluid model, presented in Figure 9(b). Of course, the time evolution of the currents is different between these two simulation cases, with the simulation with the "n-n DSMC" exhibiting lower-amplitude, shorter-lived oscillations. Nonetheless, the two simulations with different neutrals' model are seen to reach quite a similar steady state.

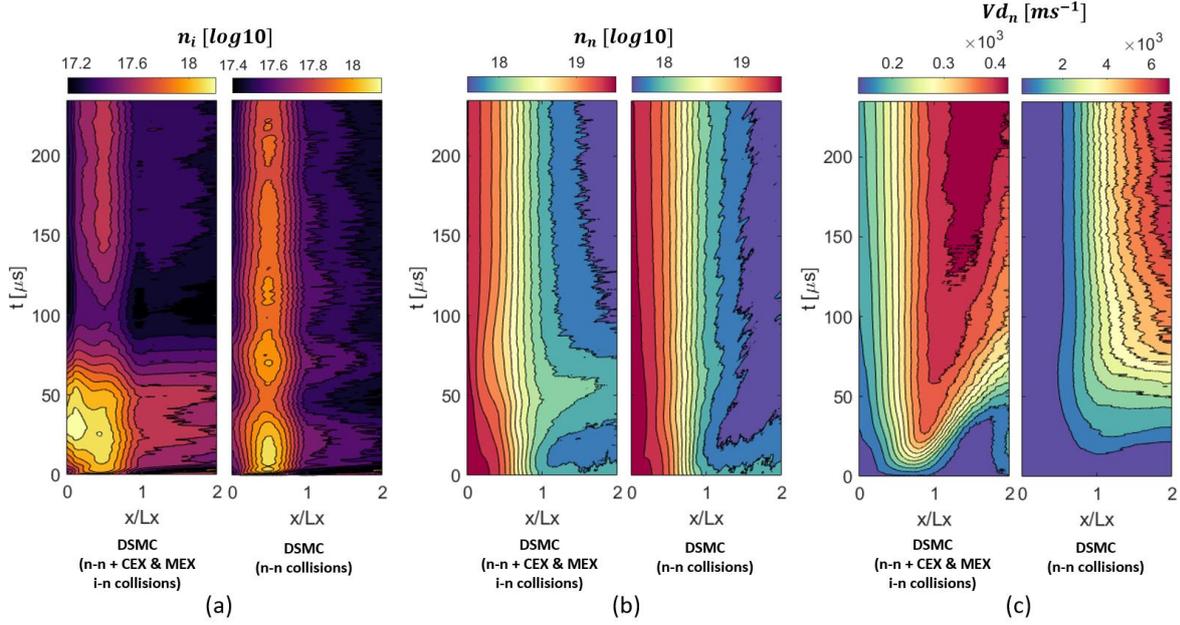

Figure 10: Time evolution of (a) ion number density, (b) neutral number density, and (c) neutral axial velocity from the 1D axial PIC simulations with the neutrals' kinetic-DSMC model.

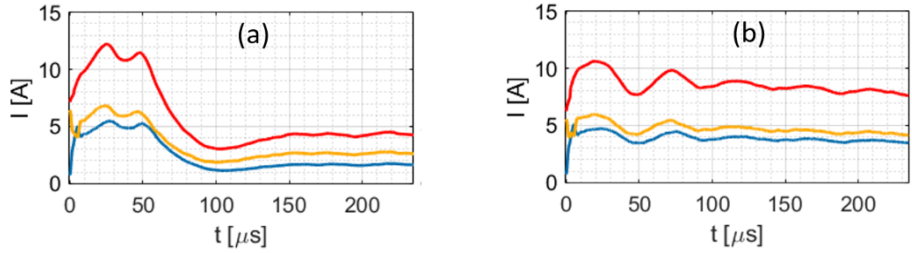

Figure 11: Time evolution of various current terms from the 1D axial simulations with neutrals' kinetic-DSMC model; (a) neutral-neutral and ion-neutral CEX and MEX collisions, (b) only neutral-neutral collisions.

The similarity of the steady-state results between the simulations with the "1D Euler" fluid and the "n-n DSMC" kinetic neutrals' model is also evident from the time-averaged axial distributions of the discharge properties shown in Figure 12. Indeed, the profiles corresponding to the "1D Euler" and the "n-n DSMC" models for the neutrals are in very good agreement. In this respect, the minor differences observed, mostly with respect to the neutral density and velocity, root in the fact that, in the neutrals' fluid model, the flow is monoenergetic whereas, in their kinetic-DSMC treatment, the particles are samples of a distribution function which leads to an inherent energy dispersion among the neutrals. Similar impacts on the discharge properties due to the neutrals' energy distribution are also reported in Ref. [21].

Another noteworthy point concerning the profiles of the "1D Euler" and the "n-n DSMC" simulation cases is that the ionization rate (Figure 12(e)) is higher in plume for the "1D Euler" compared to the "n-n DSMC". Accordingly, as the ionization results in the creation of "cold" ions, the spread in the ions' population velocity in the plume due to these cold ions is manifested as a higher time-averaged ion temperature in the plume for the "1D Euler" case (Figure 12(f)).

Regarding the time-averaged profiles of the simulation with the "n-n + i-n DMSC" collisions in Figure 12, we observe that, whereas the electric field, electron kinetic energy, and the plasma potential are consistent with the profiles from the other two simulation cases, the rest of the profiles show variations due to the ion-neutral collisions. In particular, the ion velocity shows a decrease in the plume, and the ion number density is consequently slightly increased after the exit plane. Additionally, the neutral velocity shows a significant rise from upstream the channel exit plane (dashed black line) toward the plume, which has led to a lower neutral number density inside the channel, hence, resulting in a lower ionization rate and a reduced ion density peak.

Of course, the effect of the ion-neutral collisions on the neutral population is believed to be exaggerated in the "n-n + i-n DMSC" simulation compared to what can be expected in the real world because of the lack of resolving the radial expansion of the neutrals in the simulations of this work. Such radial expansion of the neutrals lowers their density in the accelerating ion cone from the thruster, reducing the collision probability between the ions and neutrals and, hence, mitigating the momentum transfer to the neutrals.

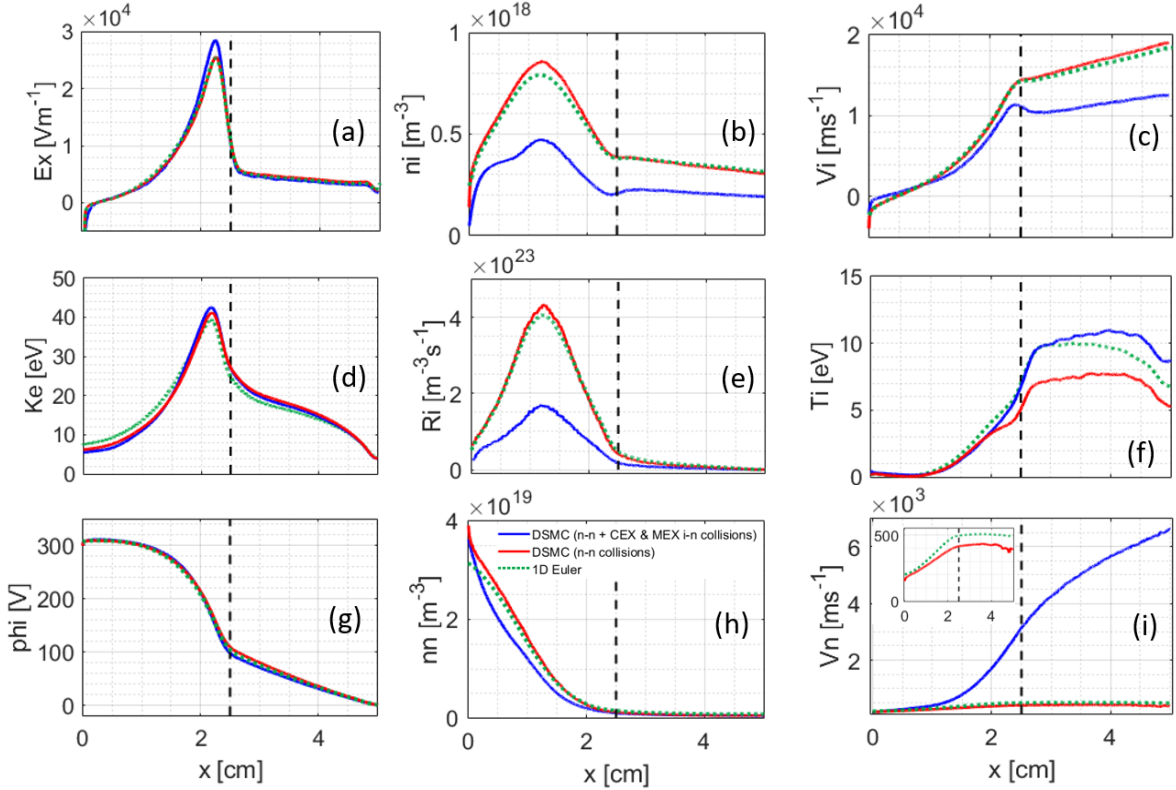

Figure 12: Time-averaged (over $100 - 230 \, \mu s$) axial profiles of the plasma and neutral properties from the 1D axial simulations with neutrals' kinetic model. The corresponding profiles from the "1D Euler" simulation are superimposed; (a) axial electric field, (b) ion number density, (c) ion axial velocity, (d) electron kinetic energy, (e) ionization rate, (f) ion temperature, (g) electric potential, (h) neutral number density, and (i) neutral axial velocity.

Comparing the profiles of the simulation in Figure 12 with those of the isothermal "1D Euler + ion momentum transfer" simulation in Figure 8, the unrealistic features in the distributions of the ion density, ion axial velocity, neutral density, and neutral velocity from the latter simulation case are not present here.

In this respect, it is recalled that, in Section 4.1, we hypothesized that the significant rise in the neutral density in the plume (Figure 8(h)) and the drop in the neutral velocity observed in Figure 8(i) are due to the occurrence of an unexpected shock wave in the neutral flow. It was also emphasized that the rise in the neutral pressure after the shock is directly reflected in the neutral density because of the isothermality assumption in the "1D Euler + ion momentum transfer" model for the neutrals.

With the above remark in mind, the neutral temperature profiles from the simulations with the kinetic "n-n + i-n DMSC" and the "full 1D Euler + ion momentum transfer" fluid model for the neutrals in Figure 13(a) illustrate that, in case the ion-neutral collisions are resolved, the neutral temperature greatly varies and, thus, the isothermality assumption can no longer be made. Therefore, it is clear that resolving the energy equation in the fluid description of the neutrals is essential in case the effect of the ion-neutral collisions on the neutral flow is to be considered in the fluid system of equations. The consequent variations in the neutral temperature, thus, prevents the formation of an unexpected shock wave that yielded unrealistic distributions of the neutral properties in the isothermal "1D Euler + ion momentum transfer" simulation.

To conclude the discussions, we refer to Figure 13(b) that shows the time-averaged axial profiles of the volumetric momentum transfer from the ions to the neutrals ($S_m$) from the simulations with the kinetic-DSMC, the isothermal "1D Euler + ion momentum transfer", and the "full 1D Euler + ion momentum transfer" neutrals' models.

Regarding this plot, it is first noted that the conspicuous difference in the $S_m$ profiles, especially in the plume region, between the "n-n + i-n DSMC" and the isothermal "1D Euler + ion momentum transfer" simulations is related to the fact that, in the "1D Euler + ion momentum transfer" case, the neutral number density in the plume was almost as large as that near the anode as a consequence of the occurred shock in that simulation. In contrast, the $S_m$ profile from the "n-n + i-n DSMC" simulation is physically reasonable since the maximum ion-neutral momentum transfer is occurring near the exit plane where the ions have been mostly accelerated and the particles' number density is still relatively high.

Second, the $S_m$ profile from the "full 1D Euler + ion momentum transfer" simulation (dotted green curve) is in close agreement with the corresponding profile from the isothermal "1D Euler + ion momentum transfer" simulation until $x \sim 2$ cm but does not exhibit a large rise in the plume. Moreover, the $S_m$ values within the channel, i.e., before the vertical dashed black line, are quite lower in the "full 1D Euler + ion momentum transfer" case compared to the $S_m$ values from the "n-n + i-n DSMC" simulation. The reason for this difference is thought to be the fact that, in the fluid description of the neutrals, the particles only have an axial velocity component whereas, in their kinetic description, the neutrals have three velocity components. Consequently, the collision of the 3V kinetic neutrals with the ions can lead to a larger change in the total momentum of the ion population.

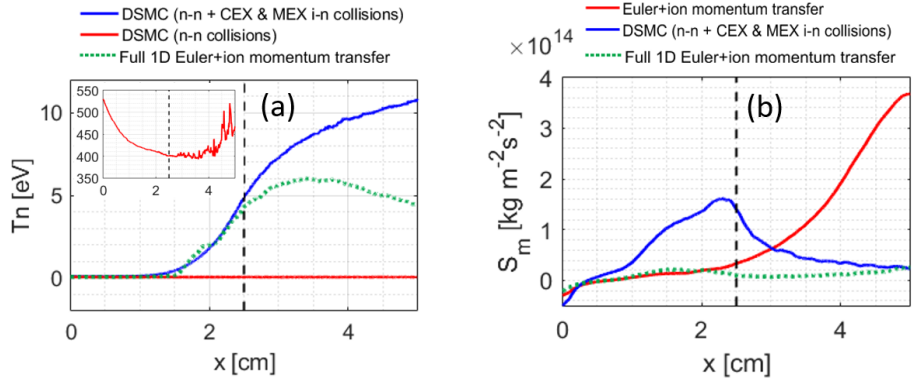

Figure 13: (a) Time-averaged axial profiles of the neutral temperature from the 1D axial simulations with the kinetic-DSMC and the "full 1D Euler + ion momentum transfer" fluid model of the neutrals, (b) time-averaged axial profiles of the ion-neutral momentum transfer term from the 1D axial simulations with the "n-n + i-n DSMC", the isothermal "1D Euler + ion momentum transfer", and the "full 1D Euler + ion momentum transfer" neutrals' models.

### 4.2.2. Results of the simulations with the collisionless neutrals' kinetic model

To assess the importance that resolving collisions between the neutrals in their kinetic description has on the dynamics of the plasma discharge, we performed a set of 1D axial simulations with the neutrals' kinetic model without any neutral-neutral collision and using three different neutral injection drift velocities. The results are shown in Figure 14 and Figure 15 in terms of the spatiotemporal evolution of the ion and neutral number density and the time evolution of the discharge current, respectively.

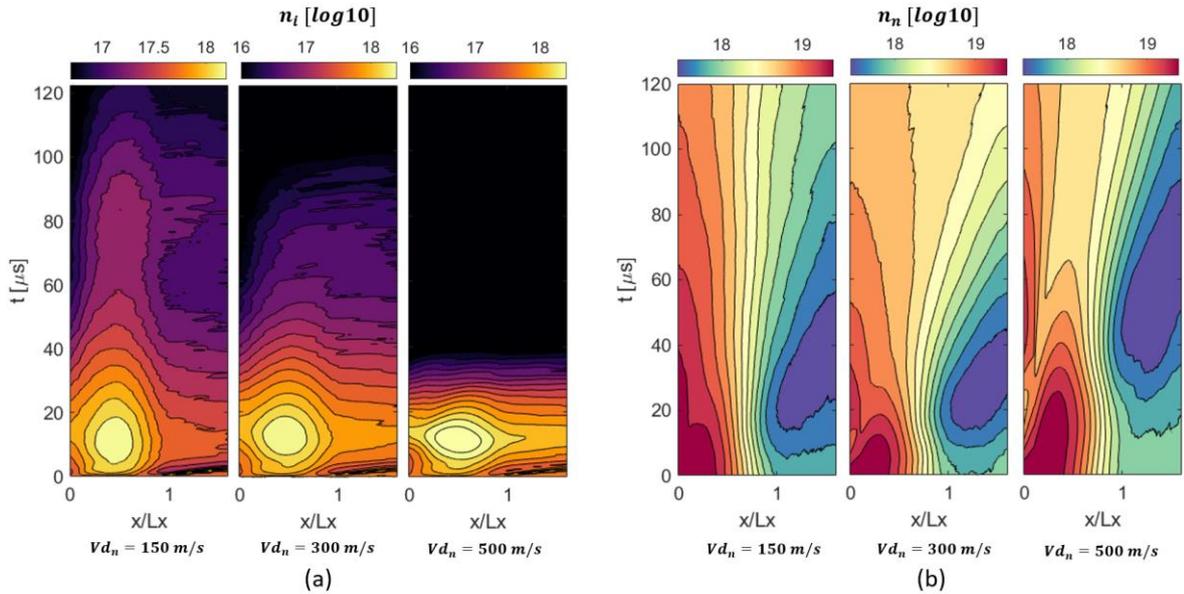

Figure 14: Time evolution of (a) ion number density and (b) neutral number density from the 1D axial simulations with the collisionless neutrals' kinetic model and various neutral injection drift velocities.

It is interesting from the plots in these figures that, regardless of the value of the neutral axial velocity, in the absence of the neutral-neutral collisions and, hence, without resolving the effect of neutral pressure, the discharge cannot be sustained and is extinguished. Nevertheless, the increase in the neutral drift velocity is observed to result in more pronounced initial transients in the discharge properties and a shorter longevity of the discharge.

Comparing the results in this section with those presented in the preceding one, particularly, in Figure 10 and Figure 11, it is evident that the neutral pressure term plays a crucial role in sustaining the plasma discharge. Indeed, in the absence of any collisions between the neutrals, the residence time of the particles within the channel seems to be insufficient to maintain the discharge following the initial transient. In this respect, it is observed from Figure 14(b) that, as the simulation progresses, the neutral density in the plume increases. This can in turn impede the effective ionization by the electrons to enable the sustainment of the discharge.

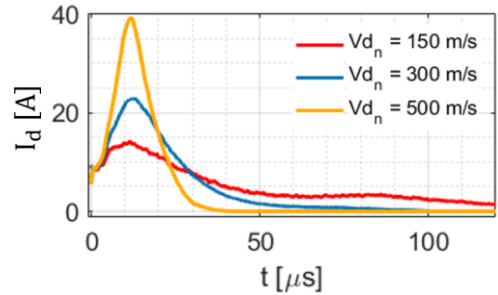

Figure 15: Time evolution of the discharge current from the 1D axial simulations with the collisionless neutrals' kinetic model and various injection drift velocities.

### 4.3. Effect of the ion recombination at the anode surface in 1D PIC simulations with various neutrals' fluid models

In this section, using kinetic 1D axial simulations with various neutrals' fluid models, we evaluate the impact of the ion recombination at the anode on the discharge behavior. In this respect, Figure 16 presents the time-averaged axial profiles of the ion number density, neutral number density, and the ionization rate as well as the time evolution of the discharge current from simulations with and without the anode ion recombination.

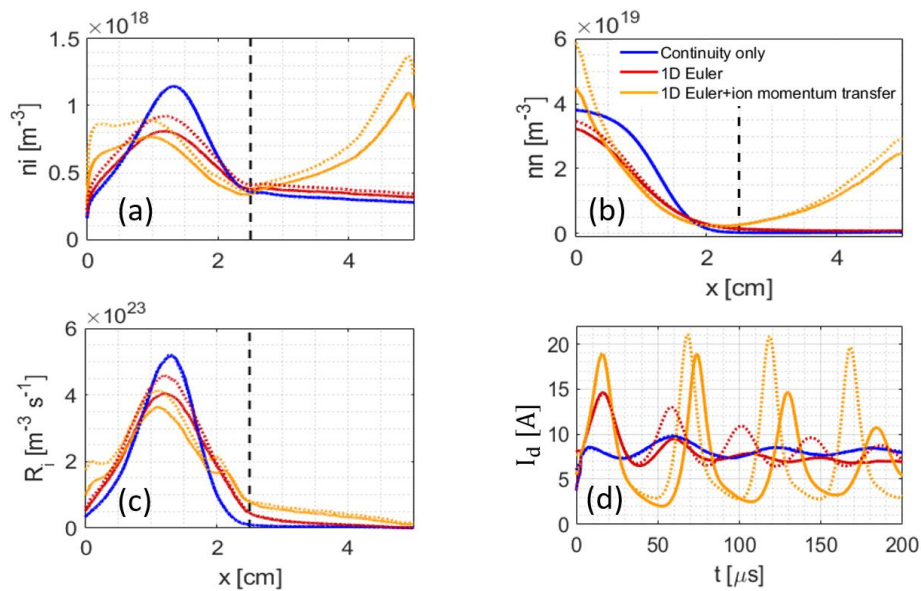

Figure 16: Time-averaged (over 200 $\mu s$) axial profiles of (a) ion number density, (b) neutral number density, and (c) ionization rate from the 1D axial simulations with different neutrals' fluid models in the presence of the anode ion recombination (dotted curves) and in its absence (solid curves). In plot (d), the corresponding time evolutions of the discharge current from various simulations are shown.

It is observed from the plots in Figure 16 that the anode ion recombination does not introduce any noticeable variation in the time-averaged properties and the discharge current oscillations in case of the "Continuity Only" model. However, in case of the "1D Euler" and the isothermal "1D Euler + ion momentum transfer" models, in both of which the variation in neutral velocity is also resolved, we see that the anode ion recombination has affected the axial distributions of the discharge properties and its temporal behavior.

Concerning the time-averaged properties, as one might have expected, the additional neutral flux due to the ion recombination at the anode has resulted in an increase in the near-anode neutral number density, which translates into a higher ionization rate and, thus, a higher ion number density in that zone. When the effect of the ion-neutral momentum exchange is also considered, we notice that, in the near-anode zone where the simulation with the isothermal "1D Euler + ion momentum transfer" model provides realistic predictions, the increase in the neutral and ion density due to the ion recombination is about 25 %.

More interestingly, we see from Figure 16(d) that not resolving the anode ion recombination yields a relative stabilization of the discharge by lowering the amplitude of the discharge current oscillations in the "1D Euler" and the isothermal "1D Euler + ion momentum transfer" simulations. In addition, the frequency of oscillations is observed to decrease in both simulation cases in the absence of the anode ion recombination.

### 4.4. Effect of the electron-wall collisions in 1D PIC simulations with the "Continuity Only" neutrals' fluid model

We evaluate in this section the effect of the electron-wall collisions on the predictions of the 1D axial simulations with the "Continuity Only" neutrals' fluid model. To this end, we have performed a simulation with no wall collisions ($\nu_{wall} = 0$) and compared its results against those from the simulations with two other values of the ad-hoc electron-wall collision frequency, namely, $\nu_{wall} = 5 \times 10^6$ and $1 \times 10^7 \ s^{-1}$. The latter value of the wall collision frequency is the one we had used in the simulations presented so far, which is also the same value used in the reference LANDMARK simulation [25]. Accordingly, since we demonstrated in Section 4.1 that our 1D axial PIC simulation with the "Continuity Only" neutrals' fluid model provides results that compare very closely with the LANDMARK, the simulations in this section can also serve as a parametric-study extension to the LANDMARK case studies [25].

Figure 17 shows the spatiotemporal maps of the ion number density from the simulations with various wall collision frequencies. In addition, for the three $\nu_{wall}$ values, Figure 18 presents the corresponding simulation results in terms of the axial distributions of the electron ($J_e$) and ion ($J_i$) current densities as well as the time variations in the discharge current.

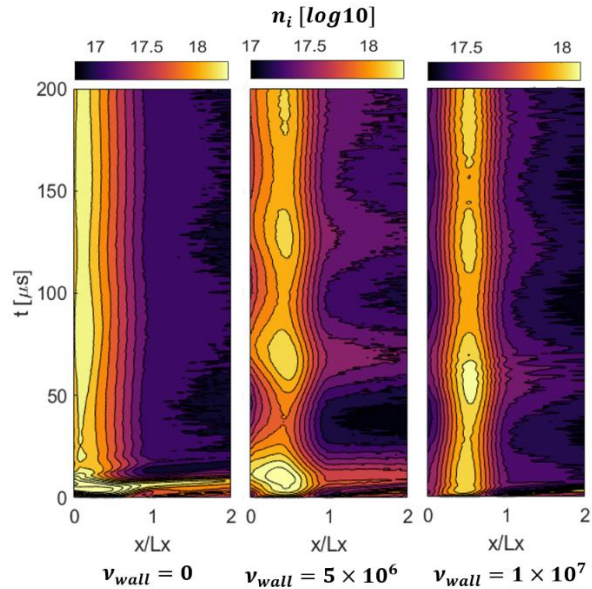

Figure 17: Time evolution of the ion number density from the 1D axial simulations with various wall collision frequencies.

It is observed in Figure 17 and Figure 18(b) that, in the absence of any wall collision, the discharge arrives at a steady state characterized by very weak oscillations after a large-amplitude initial transient in the ion number density and the discharge current. However, when electron-wall collisions are taken into account, the discharge exhibits oscillatory patterns at the system's quasi-steady state. Furthermore, increasing the frequency of the electron-wall collisions is seen from Figure 17 and Figure 18(b) to yield smaller amplitude oscillations in the discharge properties, both at the steady state and during the initial transient. Nevertheless, as the wall-induced electrons' axial mobility increases at higher wall collision frequencies, the plots (a) and (b) in Figure 18 show, respectively, that the electron current density and the mean discharge current are largest for $\nu_{wall} = 1 \times 10^7 \ s^{-1}$ compared to the other two cases.

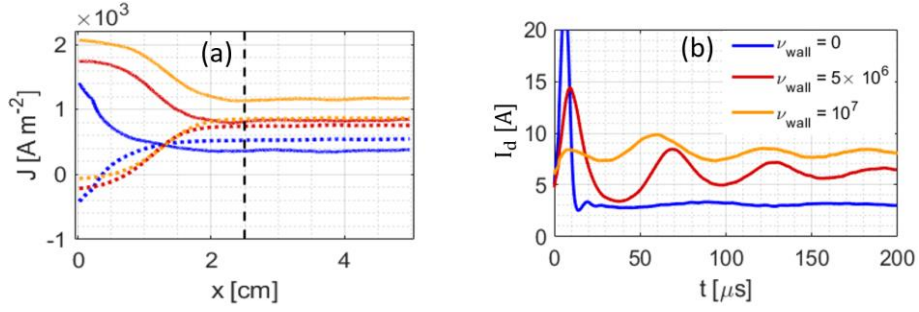

Figure 18: (a) time-averaged axial profiles of the electron (solid curves) and the ion current densities (dotted curves), and (b) the time evolution of the discharge current, from the 1D axial simulations with various wall collision frequencies.

In Figure 19, we have presented the time-averaged axial profiles of the discharge properties from the simulations with various $\nu_{wall}$ values. The most evident point from the plots in Figure 19 is that the profiles corresponding to the no-wall-collision simulation are considerably different from the profiles of the discharge properties from the other simulations. Indeed, in the absence of the wall collisions, the electron kinetic energy (Figure 19(d)) is not limited by the wall losses which has led to significant ionization and, hence, the depletion of the neutrals in the near-anode region (Figure 19(e) and (h)) with an associated large ion number density near the anode (Figure 19(b)). Moreover, the larger electron energy has resulted in a rise in the plasma potential in the near-anode zone (Figure 19(g)), which has, in turn, translated into a relatively larger electric field (Figure 19(a)) and, consequently, larger ion axial velocities (Figure 19(c)).

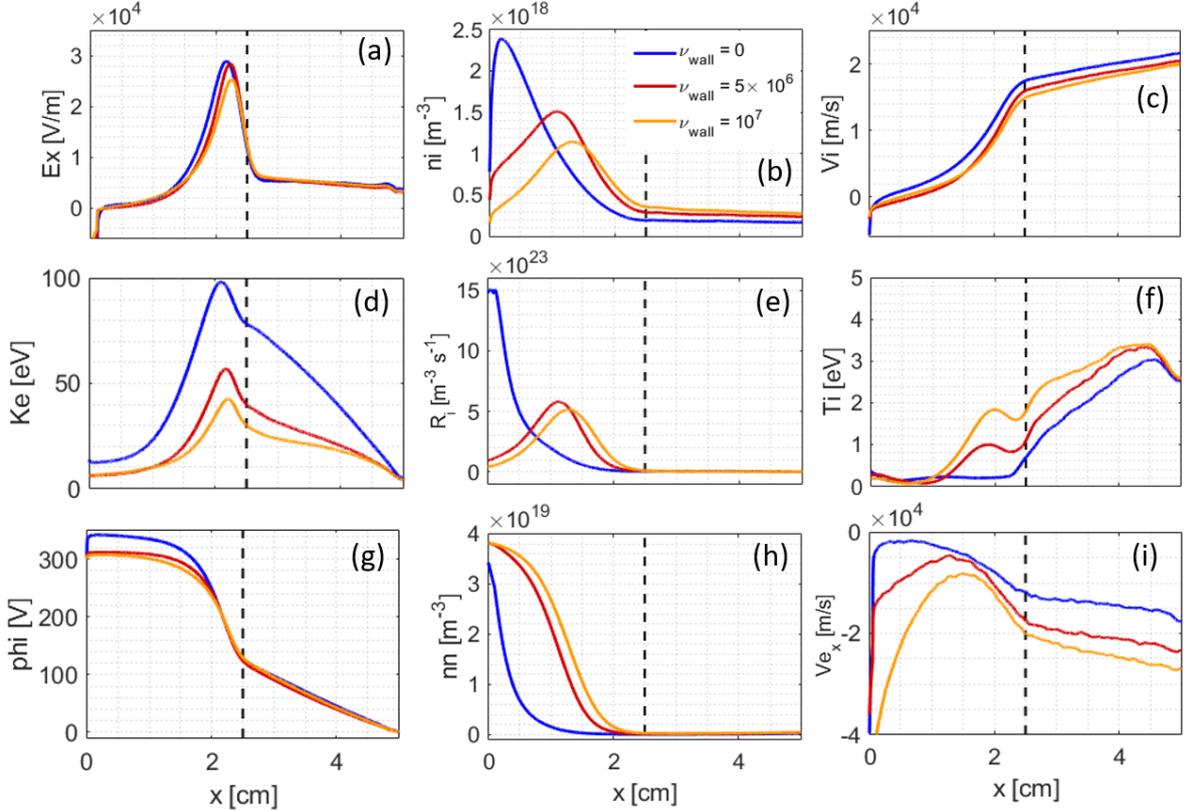

Figure 19: Time-averaged (over $200 \ \mu s$) axial profiles of the plasma and neutral properties from the 1D axial simulations with various wall collision frequencies, (a) axial electric field, (b) ion number density, (c) ion axial velocity, (d) electron kinetic energy, (e) ionization rate, (f) ion temperature, (g) electric potential, (h) neutral number density, and (i) electron axial velocity.

Taking the electron-wall collision into account in the simulations is observed in Figure 19 to result in a reduction of the predicted electron kinetic energy, which is reflected in the axial profiles of the properties mostly affected by the energy of the electron species, such as the ionization rate and the ion number density. Moreover, the increase in the axial electron drift velocity toward the anode in case of higher wall collision frequencies (Figure 19(i)) is noticed to cause a slight decrease in the slope of the plasma potential (Figure 19(g)) and, thus, the peak magnitude of the electric field (Figure 19(a)).

The observed influence of the wall collisions on the time-averaged axial distributions of the discharge properties is reflected as well in the global performance parameters presented in Table 2. The performance parameters, thrust ($T$), specific impulse ($I_{sp}$), and thrust efficiency ($\eta_T$) are calculated using the Eqs. 20 to 22. In these equations, subscript $s$ denotes a specific species, $A$ is the cross-section area, $m_s$ is the species mass, $n_s$ is the density, and $v_{sx}$ is the axial drift velocity. $\dot{m}_{out,s}$ is the species exit mass flow rate, calculated as $\sum_{ts=1}^{n} \sum_{j=1}^{N_{cell}} n_s(j,t) v_{sx}(j,t)$, $g$ is the gravitational acceleration at the sea level, and $V_a$ is the anode voltage.

The summation $\sum_{ts=1}^{n}$ in the expression for $\dot{m}_{out,s}$ and in Eq. 20 denotes the temporal average, which is taken here over the 200 $\mu s$ of the simulation time. In addition, the summation $\sum_{j=1}^{N_{cell}}$ denotes the spatial average, which, in this work, is taken over the last 100 cells of the domain.

$$T_s = \frac{A\,m_s}{n\,N_{cell}} \sum_{ts=1}^{n} \sum_{j=1}^{N_{cell}} n_s(j,t)\; v_{sx}^2(j,t)\,, \qquad \text{(Eq. 20)}$$

$$I_{sp,s} = \frac{T_s}{\dot{m}_{out,s}\,g}, \qquad \text{(Eq. 21)}$$

$$\eta_T = \frac{\Sigma_s (T_s)^2}{2\,\dot{m}_a\,I_d V_a}. \qquad \text{(Eq. 22)}$$

For the performance parameters reported in Table 2, the thrust and the specific impulse due to the neutrals as given by Eqs. 20 and 21 were found to be negligible and are, thus, neglected. Referring to Table 2, we see that the higher the wall collision frequency, the larger are the current terms, particularly the electron current. In addition, even though the thrust has increased with the wall collision frequency due to higher ion number densities in the plume (Figure 19(b)), the notable increase in the discharge current, mostly because of the enhanced electrons' axial current in the presence of the wall collisions, has led to an overall degradation of the thrust efficiency for higher wall collision frequencies.

| | 1D Axial simulations with Bohm mobility and ad-hoc wall collision model | | |
|---|---|---|---|
| | $\nu_{wall} = 0$ | $\nu_{wall} = 5 \times 10^6$ | $\nu_{wall} = 1 \times 10^7$ |
| $I_d$ (A) | 4.3 | 6.55 | 8.11 |
| $I_i$ (A) | 2.6 | 3.17 | 3.44 |
| $I_e$ (A) | 1.7 | 3.38 | 4.67 |
| $T$ (mN) | 72 | 85 | 94 |
| $I_{sp}$ (s) | 2206 | 2085 | 2031 |
| $\eta_T$ (%) | 47 | 38 | 37 |

Table 2: Global performance parameters for the simulations with various wall collision frequencies

### 4.5. Q2D axial-azimuthal simulations with different fluid models of the neutral dynamics

We report in this section the results of our reduced-order Q2D simulations in the axial-azimuthal configuration. The structuring of the discussions is overall similar to that in Section 4.1. The Q2D simulations are performed using the single-region decomposition [23], which resolves an average effect of the azimuthal instabilities on the electrons' transport [22]. We had verified in our previous publications [22][33] that this self-consistent average representation of the electrons' instability-induced cross-field mobility obviates the need for any ad-hoc transport model. Hence, the results discussed in the following are obtained without the Bohm-type transport model used for the 1D axial simulations.

The simulations, however, do include the ad-hoc model for the electron-wall collisions described in Section 2 that had also been used for the 1D axial simulations. We found that a value of the ad-hoc wall collision frequency half of that used for the 1D axial simulation, i.e., $v_{wall} = 5 \times 10^6 \ s^{-1}$, enables the Q2D simulations to arrive at a quasi-steady state. Indeed, we performed a first set of Q2D simulations with the same $v_{wall}$ value as that of the 1D axial simulations but noticed that the discharge is extinguished regardless of the neutral dynamics model. In this regard, we highlight that, since the wall collision frequency of $1 \times 10^7 \ s^{-1}$ was adopted in the LANDMARK for the axial simulation cases with a specific tuning of the Bohm transport model, it was expected that the same value may not be suitable for the Q2D simulations with a self-consistent resolution of the variations in electrons' momentum and energy due to the azimuthal waves.

Before proceeding further, we refer to Figure 20, which presents the spatiotemporal map of the azimuthal electric field fluctuations from the single-region Q2D simulation with the "1D Euler" neutrals model. It is observed that regular, small-scale wave structures have formed. This observation, which was also made for the Q2D simulations with the other fluid models of the neutrals, demonstrates that the adopted azimuthal length of the domain (1 cm) has been sufficient to capture several wavelengths of the azimuthal waves.

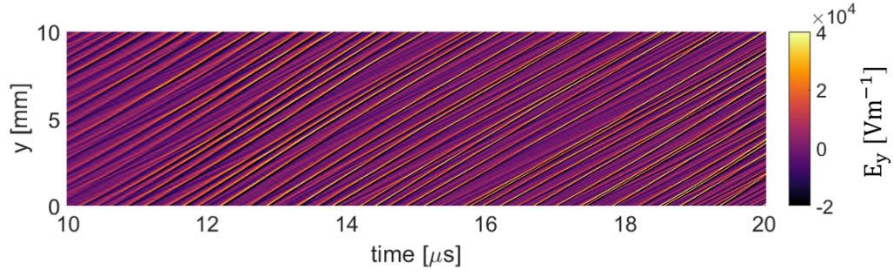

Figure 20: Spatiotemporal evolution of the azimuthal electric field fluctuations in the timeframe of 10-20 $\mu s$ from the Q2D simulation with "1D Euler" neutrals' fluid model

Figure 21 to Figure 24 show, respectively, the spatiotemporal evolutions of the discharge properties, ion number density, ionization rate, neutral density, and neutral axial velocity from various Q2D simulations. The immediate interesting observation to underline is that the single-region Q2D simulations have captured sufficient instability-induced electrons' axial mobility so that the discharge can be sustained and exhibit a periodic oscillating behavior without any ad-hoc transport model. This observation confirms the results from our previous evaluations of the reduced-order PIC scheme [22][33] but in a simulation setup that features a self-consistent description of the ionization process.

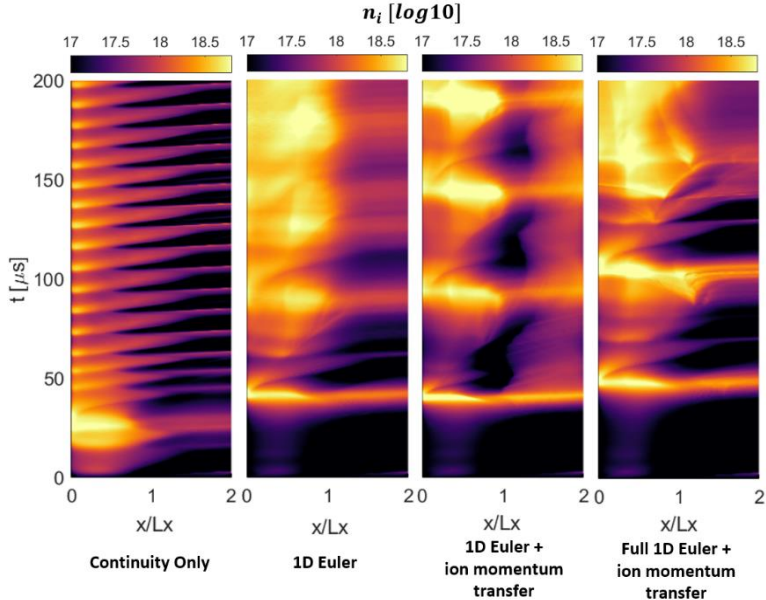

Figure 21: Spatiotemporal evolution of the ion number density from the Q2D axial-azimuthal simulations with different neutrals' fluid models.

It is also noticed from Figure 21 and Figure 22 that, compared to the 1D axial simulations (Figure 4 and Figure 5), the transient phase of the discharge takes a longer time with all models of the neutral dynamics. This is perhaps because of the time needed for the azimuthal waves to develop and for the electrons' transport to be regularized by the instability-induced processes.

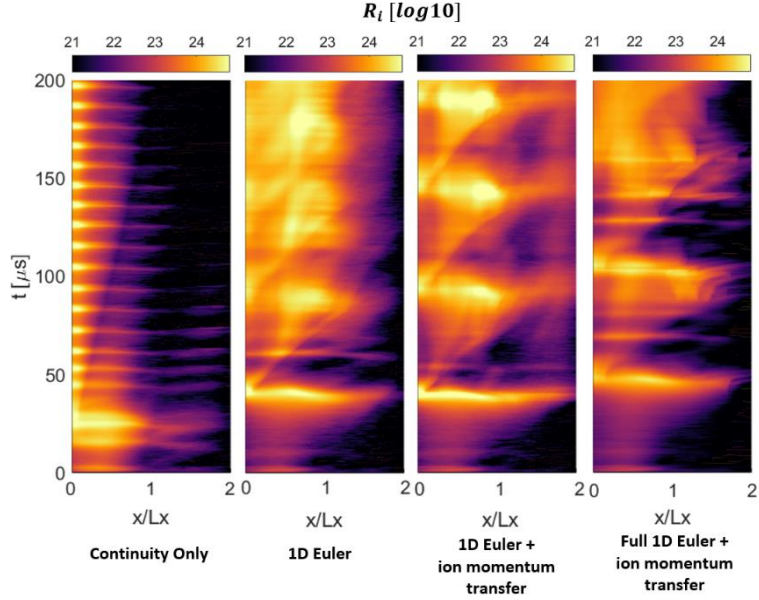

Figure 22: Spatiotemporal evolution of the ionization rate from the Q2D axial-azimuthal simulations with different neutrals' fluid models.

Comparing the spatiotemporal maps in Figure 23 and Figure 24 against those in Figure 6 and Figure 7 corresponding to the 1D axial simulations, we observe that the resolved dynamics of the neutrals is overall similar between the 1D axial and Q2D axial-azimuthal simulations for each specific neutrals' fluid model. Nevertheless, the oscillatory patterns in the neutral density and axial velocity for the "1D Euler" model are seen to be more pronounced in the axial-azimuthal simulation.

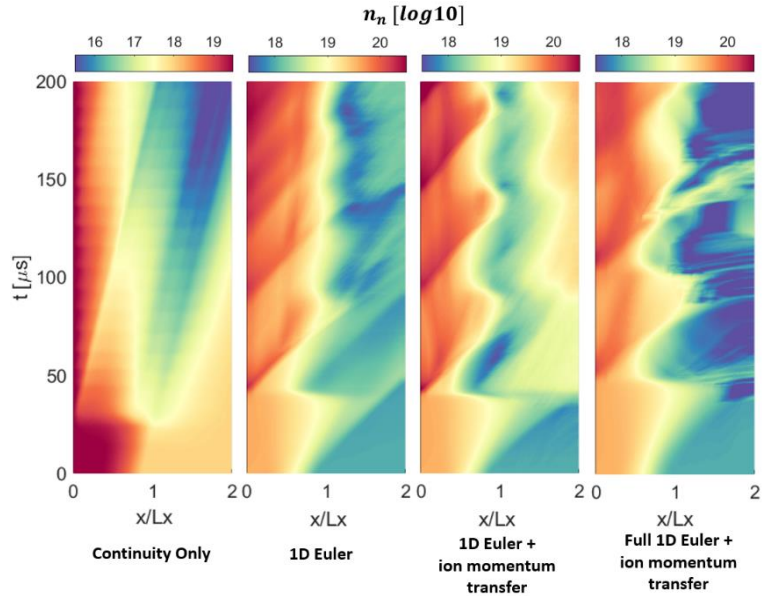

Figure 23: Spatiotemporal evolution of the neutral number density from the Q2D axial-azimuthal simulations with different neutrals' fluid models.

Another observation to point out from the plots in Figure 21 and Figure 22 is that the Q2D simulation with the "Continuity Only" neutrals' fluid model predicts a rather fast dynamics of the discharge, with the oscillations' frequency being about 100 kHz. Such a high frequency is not typically expected for the global oscillations in the discharge which are reported in the literature to be typically in the range of 5-25 kHz [2][34]. Moreover, as it is seen in Figure 25(a), these fast global oscillations are the only dominant mode in the discharge current, which points to the fact that the oscillations cannot be ascribed to the Ion Transit Time Instability (ITTI) [35]. As a result, we conclude that, in a self-consistent axial-azimuthal simulation, the "Continuity Only" model is not an appropriate choice and does not allow us to properly resolve the system's dynamics.

In fact, we observe in Figure 25(a) that, after an initial transient, the evolution of the discharge, ion and electron currents in this simulation case is characterized by low-amplitude, high-frequency oscillations, distinctly different from the evolution plots of the current terms from the simulations with the "1D Euler", isothermal "1D Euler + ion momentum transfer", and "Full 1D Euler + ion momentum transfer" neutrals' models (Figure 25(b) to (d)).

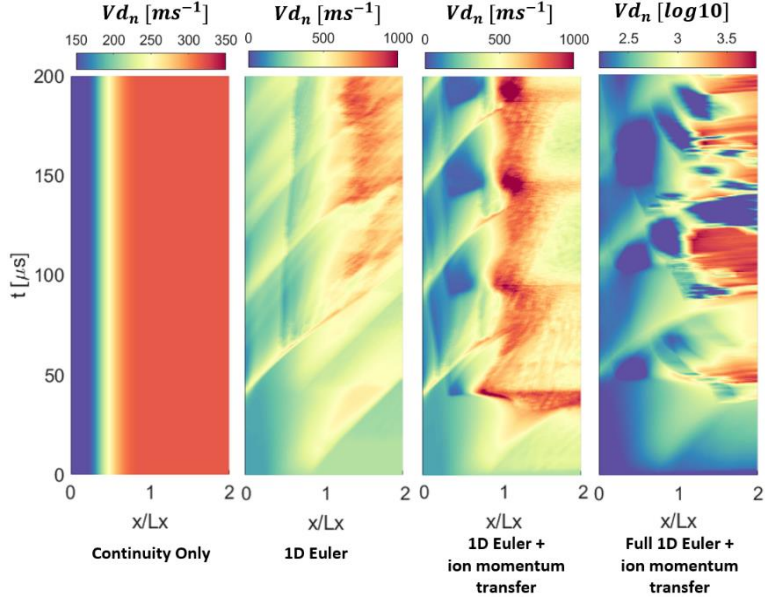

Figure 24: Spatiotemporal evolution of the neutral axial velocity from the Q2D axial-azimuthal simulations with different neutrals' fluid models.

Regarding the axial-azimuthal "1D Euler" and isothermal "1D Euler + ion momentum transfer" simulation cases, it is noticed from plots (b) and (c) in Figure 25 that the frequency of the global oscillations is very similar with the two neutrals' models (about 15 kHz), and the main difference is in the amplitude of the oscillations, which is higher from the "1D Euler + ion momentum transfer" simulation. Nevertheless, the predicted global oscillations frequency from the "Full 1D Euler + ion momentum transfer" Q2D simulation (Figure 25(d)) is about 10 kHz which is lower than that from the simulations that do not resolve the neutral's energy evolution.

As the final point regarding Figure 25, we refer to plot (e), which shows the evolution of the current terms from a 1D axial simulation with the "1D Euler" neutrals' description without a Bohm mobility model and only including the ad-hoc wall-collisions model. For this simulation case, the currents are very low and the electron current, driven only by the wall collisions, is almost zero. Comparing the current evolution trends from this case with those from the Q2D simulations, the ability of the single-region Q2D simulation in resolving sufficient electrons' transport to sustain the discharge and to allow the system to reach a quasi-steady state is clearly demonstrated.

Figure 26 presents the time-averaged axial profiles of the plasma and neutral properties from the Q2D simulations with various neutrals' models as well as from the 1D axial no-Bohm-mobility simulation with the "1D Euler" model. For the latter simulation case, whose axial profiles are illustrated as dotted black curves in the plots of Figure 26, we observe that, due to the lack of sufficient electrons' axial mobility, the plasma is unable to properly develop and, hence, the overall ion number density (Figure 26(b)) is much lower than that from the Q2D simulations. In addition, the electric potential in Figure 26(g) has a modest slope for the 1D axial no-Bohm-mobility simulation, which translates into a notably lower peak of the electric field compared to the "1D Euler", isothermal "1D Euler + ion momentum transfer" and "Full 1D Euler + ion momentum transfer" simulation cases (Figure 26(a)).

Looking at the profiles corresponding to the "Continuity Only" Q2D simulations, i.e., the blue curves in Figure 26, it is observed that the axial gradient in the electric potential is quite low in this case as well such that the slope inside the channel is almost the same as the slope for the no-Bohm-mobility simulation. Accordingly, the electric field profile exhibits a plateau near the exit plane (indicated by a vertical dashed black line) and in the near plume. Consequently, the ion axial velocity (Figure 26(c)) for the "Continuity Only" case shows a very gradual increase.

The electron kinetic energy profile in Figure 26(d) also features a plateau-like behavior around the exit plane for the "Continuity Only" case.

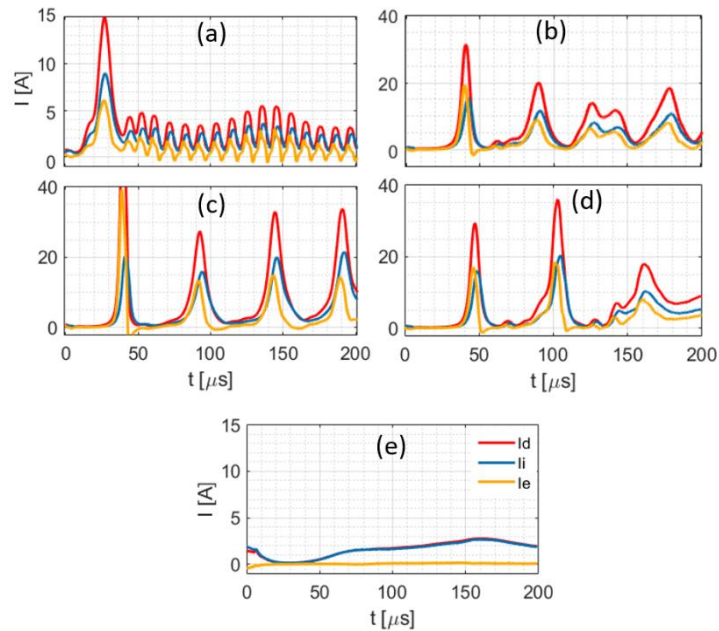

Figure 25: Time evolution of various current terms from the Q2D axial-azimuthal simulations with the neutrals' fluid models: (a) "Continuity Only", (b) "1D Euler", (c) isothermal "1D Euler + ion momentum transfer", and (d) "Full 1D Euler + ion momentum transfer" . Plot (e) shows, for reference, the time evolution of the current terms for a 1D axial simulation with no ad-hoc electron mobility.

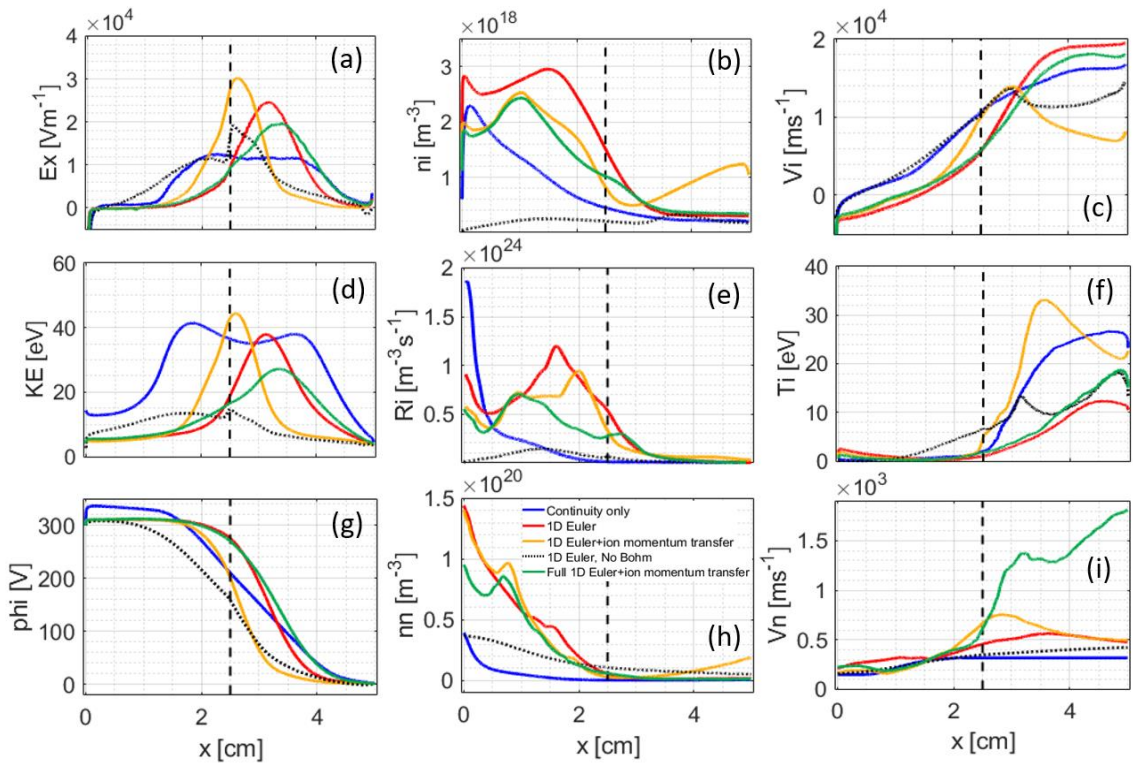

Figure 26: Time-averaged (over $200\ \mu s$) axial profiles of the plasma and neutral properties from the Q2D axial-azimuthal simulations with different neutrals' fluid models and from the 1D axial simulation with no ad-hoc mobility, (a) axial electric field, (b) ion number density, (c) ion axial velocity, (d) electron kinetic energy, (e) ionization rate, (f) ion temperature, (g) electric potential, (h) neutral number density, and (i) neutral axial velocity.

Regarding the Q2D simulations with the "1D Euler", "1D Euler + ion momentum transfer", and "Full 1D Euler + ion momentum transfer" neutrals' models, it is seen that the time-averaged profiles are quite different from the two other simulation cases discussed so far. Particularly, the peaks of the axial electric field for these cases are higher. Also, the locations of the electric field peak in the "1D Euler", "1D Euler + ion momentum transfer", and "Full 1D Euler + ion momentum transfer" axial-azimuthal simulations are shifted toward downstream compared to the electric field profiles from the corresponding 1D axial simulations in Section 4.1. Additionally, the simulation with "Full 1D Euler + ion momentum transfer" model has the most outward location of the electric field peak with the maximum electric field intensity that is lower than the simulations with "1D Euler" and isothermal "1D Euler + ion momentum transfer" models. The observations here highlight the strong influence that the selection of the neutrals' fluid model has on the distribution of the plasma properties, particularly when the electrons' mobility is self-consistently resolved.

The distribution of the ion number density from the "1D Euler" Q2D simulation has been found to be quite similar to the profile obtained from full-2D axial-azimuthal simulations of a similar simulation setup with the same neutrals' model reported in the literature [15].

Lastly, the unrealistic increase in the neutral number density and the decrease in the neutral axial velocity in the plume, explained in Sections 4.1 and 4.2 to be due to the occurrence of an unexpected shock in the isothermal "1D Euler + ion momentum transfer" case, is also visible for the Q2D simulation with this model. However, the impact of the occurred shock on the neutral and, subsequently, the ion properties is less pronounced in the Q2D simulation compared to the 1D axial one. Moreover, similar to the 1D axial case (Section 4.1), when the "Full 1D Euler + ion momentum transfer" model is used, the Q2D simulations no longer exhibit the unrealistic features associated with the profiles of the isothermal "1D Euler + ion momentum transfer" simulation in the plume.

### 4.6. Comparison of the global performance parameters from the simulations with various neutrals' models

| | | $I_d$ [A] | $I_i$ [A] | $I_e$ [A] | T [mN] | $I_{sp}$ [s] | $\eta_T$ [%] |
|---|---|---|---|---|---|---|---|
| **Q2D Axial-azimuthal** | **Continuity only** | 2.51 | 1.8 | 0.71 | 48 | 1743 | 30.5 |
| | **1D Euler** | 7.3 | 4.2 | 3.08 | 92.8 | 2046 | 39.3 |
| | **1D Euler + Ion momentum transfer** | 8.55 | 5.63 | 2.92 | 64 | 883 | 16 |
| | **Full 1D Euler + Ion momentum transfer** | 6.5 | 3.88 | 2.62 | 97 | 1815 | 48 |
| **1D Axial with Bohm (Fluid neutrals' model)** | **Continuity only** | 8.11 | 3.44 | 4.67 | 94 | 2052 | 36.3 |
| | **1D Euler** | 7.99 | 3.70 | 4.29 | 93 | 1897 | 36 |
| | **1D Euler + Ion momentum transfer** | 7.73 | 4.00 | 3.72 | 33.7 | 495 | 5 |
| | **Full 1D Euler + Ion momentum transfer** | 8.05 | 3.61 | 4.44 | 84.5 | 1733 | 29.6 |
| **1D Axial with Bohm (Kinetic-DSMC neutrals' model)** | **n-n collisions** | 8.25 | 3.74 | 4.51 | 96 | 1973 | 37.6 |
| | **n-n + CEX & MEX i-n collisions** | 4.04 | 1.54 | 2.5 | 44.4 | 1952 | 16 |
| **Reference** | **1D Euler, no Bohm** | 1.59 | 1.55 | 0.04 | 25 | 1468 | 13 |
| | **LANDMARK** | $8.15 - 8.36$ | $3.67 - 3.68$ | $4.48 - 4.67$ | --- | --- | --- |
| | **Experiment** | 4.5 | --- | --- | 83 | 1600 | 50 |

Table 3: Summary of the predicted performance parameters from the simulations carried out with various models of the neutral dynamics. Some reference values are presented for comparison. The LANDMARK figures are from Ref. [25], whereas the experimental performance values are from Ref. [36].

We present in Table 3 a comparison between the global performance parameters from the 1D axial and Q2D axial-azimuthal simulations carried out with different models of the neutral dynamics. For reference, performance values from the 1D axial no-Bohm-mobility simulation and, where available, from the LANDMARK 1D axial case, and the experiments of the SPT100 as provided in Ref. [36] are shown. The discharge, ion and electron currents are obtained using the approach of Ref. [22]. The values of thrust ($T$), specific impulse ($I_{sp}$) and thrust efficiency ($\eta_T$) reported from our simulations comprise the ions' and neutrals' contributions and are calculated using Eqs. 20-22.

The first point concerning the performance figures in Table 3 is that the values of the current terms from the "Continuity Only" 1D axial simulation with Bohm transport model are well consistent with those reported for the LANDMARK case in Ref. [25]. Moreover, the current terms from the "1D Euler" axial simulation are also quite in line with the corresponding LANDMARK values.

The second point to highlight is that the performance parameters from the 1D axial simulation with "1D Euler" neutrals' fluid model are quite aligned with those from the 1D axial simulation with kinetic "n-n DSMC" model for the neutrals.

As the third point, we observe that the predicted thrust from the Q2D "1D Euler" simulation shows an error of about 10% compared to the experimental value whereas the specific impulse has an error of about 20%. The predicted performance from the Q2D simulation with the "Full 1D Euler + ion momentum transfer" model is interestingly quite comparable with the experimental values, with the errors being, respectively, 44% in discharge current, 17% in thrust, 13% in specific impulse, and 4% in thrust efficiency. When considering the acceptability of these errors, it should be, of course, noted that the Q2D simulations lack a self-consistent description of the plasma-wall interactions, and an ad-hoc wall model had been used instead. The ad-hoc wall model was shown in Section 4.4 to affect notably the performance predictions, especially the discharge current. Consequently, the predicted performance parameters may not be readily comparable with the experimental ones. Despite this, however, the performance predictions from the "1D Euler" and "Full 1D Euler + ion momentum transfer" simulations are reasonable. In addition, comparing the performance values from these two simulation cases against those obtained from the 1D axial no-Bohm-mobility simulation, we can observe the improvement that the self-consistent resolution of the electrons' mobility in the Q2D simulation in its simplest single-region implementation has had on the performance prediction.

Fourth, in the 1D axial and Q2D simulations with the isothermal "1D Euler + ion momentum transfer" model for the neutrals, a considerable degradation of the performance is observed. The performance degradation roots in the exaggerated effect of the ion-neutral momentum exchange on the simulations with this neutrals' fluid model. Nonetheless, it is again observed that the impact of the overpredicted ion-neutral momentum transfer is less significant in the Q2D simulation. In both 1D axial and Q2D axial-azimuthal simulations, resolving the energy equation for the neutrals when considering the ion-neutral momentum transfer has obviated the issue with the unrealistic performance degradations.

Finally, in the 1D axial simulation with kinetic "n-n + i-n" DSMC model for the neutrals, the reason behind the under-predicted performance is largely the lack of resolving the neutrals' radial expansion. This was emphasized in Section 4.2 to cause a rather unrealistic enhancement of the momentum exchange between the neutral and ion species, hence, increasing the neutral axial velocity significantly which leads to a lower ionization efficiency and, subsequently, a degradation of thrust. It is, thus, concluded that the kinetic "n-n + i-n" DSMC model for the neutrals is not an appropriate choice for the PIC simulations of Hall thrusters that do not resolve the radial coordinate.

## Section 5: Conclusions

In this work, we performed an extensive study of the effects that the choice of the model to resolve the dynamics of the neutral species have on the results of the kinetic axial and axial-azimuthal particle-in-cell simulations of the plasma discharge in Hall thrusters. Overall, we observed that the model used to resolve the neutral dynamics can significantly alter the predictions of the simulations. In particular, the spatiotemporal evolution and the time-averaged profiles of the discharge properties as well as the characteristics of the breathing mode oscillations at the quasi-steady state of the system strongly depend on the neutrals' model.

We carried out 1D axial PIC simulations that featured ad-hoc models for the electrons' axial transport and the electron-wall collisions. For these simulations, we used both fluid and kinetic-DSMC models of the neutrals. Regarding the latter, two variants of the DSMC algorithm were implemented based on whether only the neutral-neutral collisions are resolved or that both the neutral-neutral and the ion-neutral collisions are accounted for in the DSMC module.

The fluid description of the neutrals was based on the 1D Euler system of equations. In this regard, we coupled the PIC simulations in this effort to various neutrals' fluid models that differ from each other in terms of the equations retained from the Euler system. The simplest fluid model used only resolved the evolution of the neutral density and was referred to as the "Continuity Only" model. The evolution of both neutral density and velocity was resolved by solving the continuity and momentum equations for isothermal neutrals using a subset of the Euler conservation equations that we simply called the "1D Euler" model. Two additional fluid models of neutral atoms were used, in both of which the ion-neutral momentum transfer had been considered as an additional source term in the equations. However, one model, named "Full 1D Euler + ion momentum transfer", included continuity, momentum, and energy equation for the neutrals whereas the other model resolved only continuity and momentum equations for the isothermal neutral atoms, hence the name "1D Euler + ion momentum transfer".

From the axial PIC simulations, we highlighted that the time evolution of the plasma properties and their time-averaged distributions notably depend on the model adopted to describe the dynamics of the neutral properties. In this regard, three major outcomes of the axial simulations were that: (1) the "1D Euler" fluid model for the neutrals is greatly representative of their kinetic description with neutral-neutral DSMC collisions such that the predictions of an axial PIC simulation incorporating either model are highly similar in terms of the time-averaged plasma properties and the global performance parameters; (2) in the presence of the ion-neutral collisions, the temperature of the neutral atoms rise significantly because of the energy exchange between the ions and the neutrals, which implies that the isothermal "1D Euler + ion momentum transfer" model cannot properly resolve the variations in the neutral flow properties. (3) the "n-n + i-n DSMC" model that accounts for the ion-neutral collisions is not appropriate choice for the PIC simulations that do not resolve the radial coordinate. This is because the lack of capturing neutrals' expansion along the radial direction leads to an overestimation of the ion-neutral collision frequency, hence, affecting in an exaggerated manner the neutrals' and ions' properties in the plume.

Using the axial PIC simulations with different fluid descriptions of the neutrals, we also assessed the impact of the ion recombination at the anode, observing that, this phenomenon does not have any noticeable influence on the simulation's results in the case of "Continuity Only" model. However, for the "1D Euler" and "1D Euler + ion momentum transfer" models, the presence of the anode ion recombination was seen to yield an increase in the near-anode neutral and plasma density as well as an increase in the frequency and amplitude of the discharge current oscillations.

The influence of the electron-wall collisions was additionally investigated in the 1D axial simulations with the "Continuity Only" model. It was noticed that the wall collisions enhance the electrons' axial mobility and limit their kinetic energy, thus affecting the spatiotemporal evolution of the discharge and the state of the plasma at the steady state. Moreover, in terms of the performance parameters, the wall collisions were observed to result in an increased discharge current and thrust while a degradation in the specific impulse and thrust efficiency.

In the axial-azimuthal configuration, we performed reduced-order single-region Q2D simulations with the same set of neutrals' fluid models as that of the axial simulations. The Q2D simulations self-consistently resolved the electrons' axial mobility but included an ad-hoc description of the electron-wall collisions. These simulations underlined that the axially averaged instability-induced electrons' mobility captured by the first-order single-region approximation of the axial-azimuthal problem enables the discharge to be sustained and allows for the development of periodic oscillations at the quasi-steady state in the case of all neutrals' fluid models.

Nonetheless, the simulation with the "Continuity Only" model was observed to predict a fast global dynamics of the discharge with a frequency of about 100 kHz and peculiar time-averaged distributions of the plasma properties. It was, thus, concluded that, in a simulation with self-consistent description of electrons' mobility, the "Continuity Only" model is not a suitable choice.

In contrast, the "1D Euler" and "Full 1D Euler + ion momentum transfer" Q2D simulations were observed to provide reasonable predictions in terms of the plasma properties' profiles and the global performance parameters, particularly, the thrust, the specific impulse, and the thrust efficiency.

Finally, the unrealistic predictions of the isothermal "1D Euler + ion momentum transfer" model concerning the neutrals' properties in the plume were seen to be remedied by resolving the energy equation for the neutrals through the "Full 1D Euler + ion momentum transfer" model in both the 1D axial and the Q2D axial-azimuthal simulations.

It is noteworthy that, even though not presented in this paper, our preliminary Q2D axial-azimuthal simulations with a kinetic "n-n DSMC" model of the neutrals has showed that the simulation results seem to be consistent with those obtained from the fluid "1D Euler" Q2D simulation reported in this article.

## Appendix: Benchmarks for the neutrals' fluid model

### 1. Sod shock tube problem

We carried out a simulation of the famous Sod shock tube problem [37] to verify the basic implementation of the neutrals' fluid model. The problem is described by a source-free 1D Euler system of equations comprising the continuity, momentum, and energy conservation equations.

The simulation domain, shown in Figure 27, consists of a tube with a high- and a low-pressure region separated by a diaphragm at the middle. The properties of the gas are uniform in each region and are given as initial conditions. At $t_0$, the diaphragm is removed and the gas from the high-pressure region quickly expands into the low-pressure region. According to the theory, this is accompanied by the occurrence of an expansion fan and a shock wave. In between, a contact discontinuity exists that separates the regions of different entropy (or total energy). Noting the presence of the discontinuities in the problem, the use of the HLLC Reimann solver for the inter-cell fluxes is warranted since its ability to properly resolve the discontinuities in the solution is one of its strong merits.

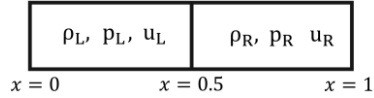

$$x = 0 \qquad x = 0.5 \qquad x = 1$$

Figure 27: Simulation domain for the Sod shock tube benchmark

In order to perform the simulation, a domain length of 1 m, a total simulation time of 0.2 s, and two different values of cell size were chosen. A simulation with $N = 200$ cells is performed to verify the accuracy of the results against a high-resolution solution given by a simulation with 8000 number of cells, which effectively amounts to the analytical solution. Dirichlet boundary conditions were applied to both ends of the domain. The CFL criterion of 0.5 is considered, and the initial conditions of Eq. A.1 were used:

$$\begin{pmatrix} \rho_L \\ p_L \\ u_L \end{pmatrix} = \begin{pmatrix} 1.0 \\ 1.0 \\ 0.0 \end{pmatrix} \text{ for } x < 0.5, \quad \text{and} \quad \begin{pmatrix} \rho_R \\ p_R \\ u_R \end{pmatrix} = \begin{pmatrix} 0.125 \\ 0.1 \\ 0.0 \end{pmatrix} \text{ for } x > 0.5. \qquad \text{(Eq. A. 1)}$$

Figure 28 presents the results of the simulations with the two values of computational cells. It is observed that the approximate solution with 200 cells is in great agreement with the high-resolution solution with 8000 cells which is exactly coincident with the steady-state solution of the problem. Only minor numerical oscillations can be noticed around the location of the discontinuities for the 200-cell simulation case. In addition, the variations in the gas properties due to the expansion fan and due to the contact and shock discontinuities are correctly captured. All these observations confirm the accurate implementation of the neutrals' fluid model.

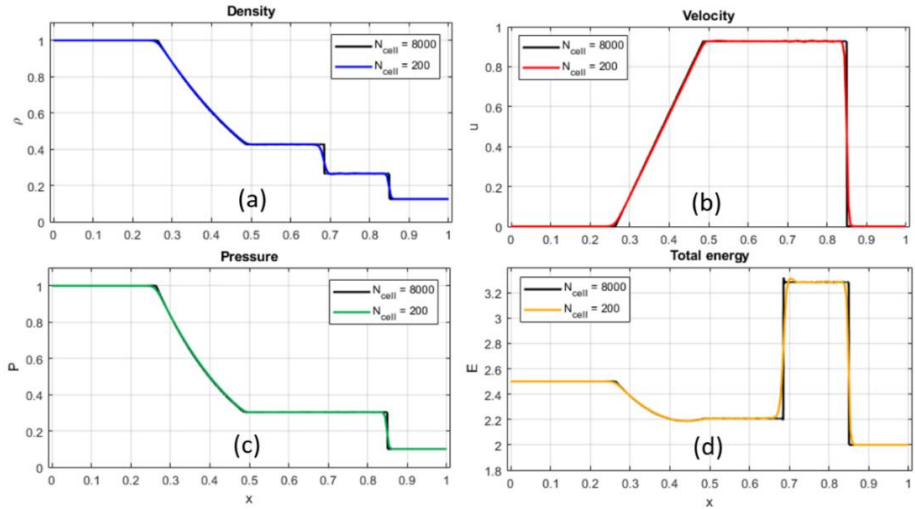

Figure 28: Profiles of the neutrals' properties from the neutrals' fluid model at the end of the Sod test problem simulation; (a) density ($\rho$), (b) velocity ($u$), (c) pressure ($p$), and (d) total energy ($E$).

## 2. Simplified Hall thruster problem for isothermal neutrals

This test, which is inspired from Ref. [15], was performed to assess the accuracy of the neutrals' fluid model, described in Section 3.1, in a simulation case representative of a Hall thruster. We consider a 1D axial domain with the length of 4 cm. A Gaussian ionization source is imposed as shown in Figure 29. The system of governing equations corresponds to that given by Eqs. 2-4, and the boundary conditions are the same as those in Eqs. 8 and 9-11 for the cathode and the anode sides of the domain, respectively. We used three variants of the Euler system of equations for the verification here; two identical to the "Continuity Only" and the "1D Euler" models introduced in Section 3.1, and the third representing the full Euler system (continuity, momentum, and energy) with a specialized energy source term ($\frac{E-E_0}{\tau}$), adopted from Ref. [15], to ensure that the neutrals will remain isothermal. In the relation for this source term, $E$ is the neutrals' specific total energy, $E_0$ is the specific total energy calculated using the initial neutrals' temperature (640 K), and $\tau$ is a relaxation factor, the value of which was set at $10^{-8}$ in accordance with Ref. [15].

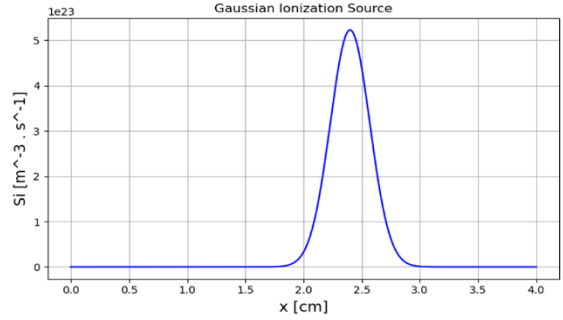

Figure 29: Profile of the imposed ionization source for the Hall thruster test problem

The initial neutral density and velocity profiles are shown as dashed blue lines in Figure 30. A mass flow rate of 5 mg/s is considered with an initial neutrals' velocity of 200 m/s. The simulation was run for $2\ ms$, using 200 cells, and a CFL criterion of 0.5. The profiles of the neutrals' properties, density, velocity, and pressure, at the system's steady state are shown in the plots (a) to (c) of Figure 30. It is seen from these plots that, for the case of isothermal neutrals, the "1D Euler" model provides results that are indistinguishable from those obtained by solving the full Euler system.

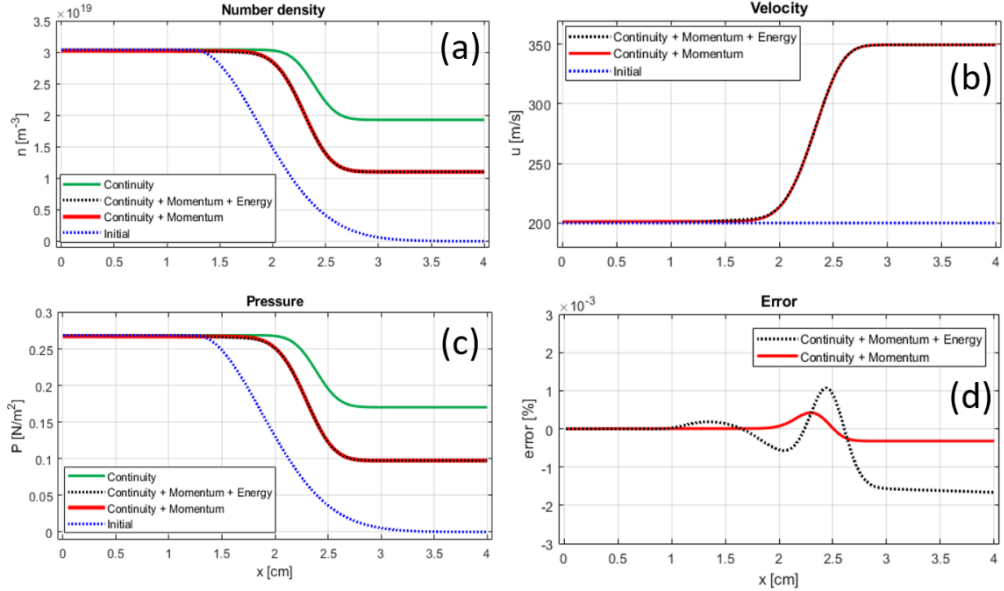

Figure 30: Verification results of the neutrals' fluid model in the simplified Hall thruster problem; profiles of various neutral properties are shown at the beginning and the end of the simulation time; (a) neutral density, (b) axial velocity, (c) neutral pressure ($p = nk_BT$). In plot (d), the error profile, calculated according to Eq. A.3, is illustrated.

To define an error figure for the results, we used the characteristic relation given by Eq. A.2, which is obtained by substituting the continuity and the equation of state for isothermal neutrals into the momentum equation and integrating the resulting ODE.

$$\frac{1}{2}u^2 + \frac{k_B T}{M}\ln(\rho) = C; \quad \text{with } C = \frac{1}{2}u_0^2 + \frac{k_B T_0}{M}ln(\rho_0). \tag{Eq. A. 2}$$

The constant $C$ is determined at the anode boundary, and the local values of neutrals' density and velocity in each cell must satisfy Eq. A.2. Thus, the error term, whose distribution is shown in Figure 30(d), is defined as

$$error = \frac{\frac{1}{2}u^2 + \frac{K_B T}{M}\ln(\rho) - C}{C}. \tag{Eq. A. 3}$$

It is evident from Figure 30(d) that the error of the solutions from the "1D Euler" and the isothermal full-Euler-system models is on the order of $10^{-3}$ %, which is highly acceptable.

**Acknowledgments**:

The present research is carried out within the framework of the project "Advanced Space Propulsion for Innovative Realization of space Exploration (ASPIRE)". ASPIRE has received funding from the European Union's Horizon 2020 Innovation Programme under the Grant Agreement No. 101004366. The views expressed herein can in no way be taken as to reflect an official opinion of the Commission of the European Union.

The authors also gratefully acknowledge the computational resources and support provided by the Imperial College Research Computing Service (http://doi.org/10.14469/hpc/2232).

**Data Availability Statement**:

The data that support the findings of this study are available from the corresponding author upon reasonable request.